# The proportional scaling of mRNA and ribosome concentrations controls eukaryotic cell growth

Xin Gao[1], Michael Lanz[1,2], Rosslyn Grosely[3], Jonas Cremer[1], Joseph Puglisi[3], and Jan M. Skotheim[1,2,*]

[1]Department of Biology, Stanford University, Stanford, California, USA

[2]Chan Zuckerberg Biohub, San Francisco, California, USA

[3]Department of Structural Biology, Stanford University School of Medicine, Stanford, California, USA

[*]Address correspondence to skotheim@stanford.edu

## ABSTRACT

Cell growth underlies nearly all eukaryotic physiology, yet its quantitative principles remain unclear. Using single-molecule ribosome tracking, spike-in RNA sequencing, and quantitative proteomics across 15 nutrient-limited conditions in budding yeast, we define how growth is controlled in the budding yeast *Saccharomyces cerevisiae*. Ribosome concentration scales linearly with growth rate, while peptide elongation speed remains constant at ~9 amino acids $s^{-1}$. While elongation is not a regulatory lever, total mRNA concentration increases proportionally with ribosomes to accelerate growth. A simple kinetic model of mRNA–ribosome binding accurately predicts the fraction of active ribosomes, growth rate, and responses to transcriptional or size perturbations. Consistent with this model, transient inhibition of mRNA degradation boosts growth by elevating mRNA concentration. These results reveal that eukaryotic cells accelerate proliferation primarily by proportionally scaling mRNA and ribosome abundance, establishing a quantitative framework for understanding eukaryotic biosynthesis.

## INTRODUCTION

Cells continuously tune their biosynthetic capacity in response to nutrient availability. A hallmark of this coordination is the increase in ribosome abundance as growth accelerates, which has been observed in diverse species including bacteria and yeast [1–10]. Elegant quantitative models developed for the bacteria *E. coli* describe growth as a mass-action process in which charged tRNAs collide with mRNA-bound ribosomes [11–15]. By balancing the proteome fractions devoted to metabolism (which supplies charged tRNAs) and translation (ribosomes which consume them), these models predict how ribosome content, active-ribosome fraction, and peptide-elongation rate change with growth rate. While these quantitative models developed for *E. coli* have proven useful, there is little in these models that is specific to bacterial physiology. This raises the question whether the same principles apply to eukaryotes [16]. Here, inspired by models of bacterial growth physiology, we aim to identify the fundamental principles governing eukaryotic growth by examining the budding yeast *S. cerevisiae*.

In *S. cerevisiae*, it is currently unclear which molecular components are limiting for cell growth. Proteome surveys report that ribosome content rises with growth rate [1,3–7], but indirect proxies such as polysome profiles suggest that a sizable fraction of ribosomes may remain inactive [4,17]. Furthermore, direct measurements of individual ribosome translation rates are limited to two-condition comparisons in previous studies, with conflicting results [17–21]. Moreover, ribosomes are not the only expensive biosynthetic investment and metabolic enzymes, tRNAs, or mRNAs could also limit growth. In *E. coli*, mRNA concentrations increase when cells grow faster [12,13,22], but it is unclear whether yeast behave similarly because the scaling of the total mRNA concentration with growth rate has never been measured directly [23–25].

Here, we measure every variable needed to construct a model of eukaryotic cell growth. Single-molecule tracking of Halo-tagged ribosomal proteins quantifies the fraction of actively translating ribosomes across 15 carbon- and nitrogen-limited conditions, while spike-in RNA-seq, tRNA RT–qPCR, and quantitative proteomics yield absolute mRNA, tRNA, and ribosome concentrations, respectively. We find that (i) the active ribosome fraction increases with growth rate; (ii) the peptide-elongation rate stays essentially constant at ~9 aa $s^{-1}$; and (iii) the total mRNA concentration increases in proportion to the growth rate. These patterns contradict canonical bacterial models but align with a new kinetic framework in which mRNA–ribosome mass-action kinetics, rather than charged tRNA, limits protein synthesis. Yeast therefore accelerates growth by up-regulating both ribosomes and mRNA, which establishes a new quantitative paradigm for eukaryotic growth physiology.

## RESULTS

To determine how yeast grow at different rates in different environments, we sought to measure the compositional changes of the major building blocks of biosynthesis, including ribosomes, tRNA, and mRNA. To do this, we started by measuring ribosome concentrations. We grew *S. cerevisiae* strains on 15 different media (12 carbon sources and 3 different nitrogen sources) and performed mass spectrometry (see **Methods,**). Consistent with previous results [3–7], we found a linear increase in the fraction of the proteome devoted to ribosomal proteins as the growth rate increases (**Fig. 1a; Extended Data Fig. 1a**). In parallel, total RNA concentration increased linearly with growth rate, mirroring the ribosomal protein fraction of the proteome (**Extended Data Fig. 1b**). Moreover, as observed for the *E. coli* bacteria [2,14,10], we found a decrease in the proteome fraction allocated to metabolic pathways as growth rate increases, reflecting a trade-off between biosynthetic and metabolic resource allocation (**Extended Data Fig. 1c**).

### The fraction of actively translating ribosomes increases with growth rate

That the concentration of ribosomes increases linearly with the growth rate suggests that this is the limiting factor for growth, *i.e.*, more ribosomes drive proportionally faster growth. However, some work suggested that there may be a pool of inactive ribosomes that complicates this simple model [4,17]. This hypothesis was supported by a recent study that estimated the fraction of active ribosomes using polysome profiling under two growth conditions [4]. However, polysome profiles are difficult to quantify and treating the proportion of polysomes as a proxy for the fraction of active ribosomes may be problematic because monosomes could also contribute to translation [26,27]. We therefore sought an alternative method to directly measure the active ribosome fraction in a variety of growth conditions.

During peptide elongation, ribosomes bound to mRNA are characterized by a slow diffusion coefficient due to the large size of mRNA, whereas inactive ribosomes diffuse freely in the cytoplasm and exhibit a high diffusion coefficient [28]. To calculate the diffusion coefficient of ribosomes and then quantify the active ribosome fraction in living yeast cells, we employed single-molecule tracking [29] (**Fig. 1b**). To track individual ribosomes in living yeast, we first deleted the membrane transporter gene *PDR5* to retain fluorescent dyes inside the cells. We then fused the Halo tag to the C-terminus of the ribosomal subunits Rps4b, Rps9a, Rpl26a, and Rpl29 [30]. Among these, *RPS9A-Halo* and *RPL26A-Halo* strains maintain growth rates similar to wild-type (**Extended Data Fig. 1d-f**). We therefore decided to use the strain expressing *RPS9A-Halo* for further analysis. Briefly, cells were labeled with the PA549 dye before we performed live-cell photoactivated localization microscopy (PALM) for over 50 cells for each condition (**Fig 1c; Extended Data Fig. 1g;** see **Methods**). The diffusion coefficient of a single ribosome $D_a$ can be calculated from the stepwise mean-squared displacement (MSD) of its trajectory. To determine the active ribosome fraction, we fit the distribution of ribosome diffusion coefficients in a population using a Gaussian mixture model composed of a slow-diffusing active fraction and a fast-diffusing inactive fraction (**Fig. 1d**).

To validate our approach, we treated cells with the translation initiation inhibitor lactimidomycin (LMT), which stalls ribosomes at the start codon and allows already elongating ribosomes to run off the mRNA. In glucose-grown cells, LMT treatment reduced the fraction of slowly diffusing ribosomes from 70% to 35% (**Fig. 1e**). Our data are consistent with the observation that mRNA concentration is approximately one-third that of the ribosome concentration [31,32]. This is because in our experiment, we expect that sometime after LMT treatment, the actively translating ribosomes would all have run off their mRNA. There would also be sufficient time for unbound ribosomes to be loaded onto the start codon of each mRNA, where it would stall so that each mRNA would then be bound to a single ribosome. These results suggest that our single-molecule tracking–based method accurately reflects the fraction of active ribosomes bound to mRNA.

After establishing our method to measure the active ribosome fraction, we sought to determine how this fraction changes in different growth conditions. To do this, we examined yeast cells growing in 12 different carbon sources and 3 different nitrogen-limited conditions [33] (**Fig. 2a-b**). The average diffusion coefficients were similar across all these growth conditions and consistent with theoretical estimates using the Stokes-Einstein equation applied to a typical 80S monosome or a 40S ribosome subunit [34,35] (**Fig. 2c; Extended Data Fig. 1h**). Moreover, our results indicate that the active ribosome fraction is positively correlated with growth rate (**Fig. 2d**). The slower cells grow, the larger the fraction of inactive ribosomes. To confirm this observation, we repeated these experiments with a strain expressing *RPL26A-Halo* to label the 60S ribosomal subunit, which can remain associated with the endoplasmic reticulum (ER) after the translation of some proteins [36]. Diffusion measurements of *RPL26A-Halo* revealed a similar increase in the active ribosome fraction with growth rate as for the 40S subunit (**Extended Data Fig. 2a-b**). We note that our estimates of the active fraction may be slightly large due to some ribosomes being associated with compartments like the ER, which can reduce their mobility even if they are not actively translating [37,38]. To confirm our results, we therefore performed polysome profiling on cells grown in four conditions. The polysome percentage rose at higher growth rates, while the 40S and 60S peaks declined, reflecting a higher proportion of ribosomes in an active state (**Fig. 2e-f; Extended Data Fig. 2c**).

Thus, we find that the fraction of the proteome that is active ribosomes increases in proportion to the growth rate, while the fraction of the proteome that is inactive ribosomes is slightly decreased from 10% to 8% (**Fig. 2g**). In other words, as yeast grow faster, they engage proportionally more ribosomes in translation while there remains a substantial inactive fraction of ribosomes.

### Global peptide elongation rate is independent of growth rate

After estimating the active ribosome fraction, we can now estimate the average peptide elongation rate of a translating ribosome, a predicted output of several growth models. We neglect protein degradation in our analysis because the vast majority of the proteome is stable on the time scale of a cell division cycle [39,40]. The peptide elongation rate, $c_p$, is related to the average protein mass per cell, $m_{prot}$, the active ribosomes per cell $N_{ribo}^{active}$, the population doubling time $T_d$ by the following equation [20]:

$$c_p = \frac{m_{prot} \ln 2}{N_{ribo}^{active} T_d}.$$

If the ribosome fraction in the proteome is $\phi_r$, the fraction of ribosomes that is active is $\phi_a$, and the average protein mass of a single ribosome is $m_{rb}$, then the average ribosomal protein mass in a cell is

$$m_{prot} \phi_r = N_{ribo} m_{rb} = \frac{N_{ribo}^{active} m_{rb}}{\phi_a}.$$

Substituting $m_{prot}$ in the equation for the peptide elongation rate yields:

$$c_p = \frac{m_{rb} \ln 2}{\phi_a \phi_r T_d}.$$

Examining the peptide elongation rate in different growth conditions revealed that it is nearly independent of growth rate (~9 amino acids per second; **Fig. 3a**). In contrast, previous studies estimated elongation rates assuming that all ribosomes were active. If we make this erroneous assumption, the average elongation rate increases with growth rate, consistent with earlier reports [17–21,41] (**Extended Data Fig. 3a**). This result is consistent with one study using pulse-chain radioactive labeling technology [17], but not another that analyzed only three small molecular weight proteins [18]. In addition, we examined whether translation efficiencies (ribosome footprints/mRNA reads for a given gene, see **Methods**) change with growth rate [42,43]. By combining ribosome footprints and RNA-seq, we found that translation efficiencies across genes span approximately two orders of magnitude. Despite the large-scale remodeling of the transcriptome in different carbon sources, the translation efficiency of individual genes remained largely stable across conditions (**Fig. 3b-c; Extended Data Fig. 3b-c**).

To validate our method for calculating peptide elongation rates, we treated yeast cells growing in four different carbon sources with the translation inhibitor cycloheximide (CHX). CHX treatment decreased the growth rate in all conditions. Compared to untreated cells, CHX-treated cells exhibited significant proteomic alterations, including to the ribosome concentration we need to calculate the peptide elongation rate (**Fig. 3d-e**; **Extended Data Fig. 3d-e**). The peptide elongation rate in CHX-treated cells was significantly reduced, consistent with expectations (**Fig. 3f**). To further test the notion that the average peptide elongation rate was similar across conditions, we compared the onset of luminescence from reporters containing variable-length coding sequences inserted upstream of the *Nluc* luciferase gene in two different growth conditions [44,45]. Consistent with our other data, we observed no significant differences in the luciferase onset time in the two growth conditions (**Extended Data Fig. 3f-g**).

That the average peptide elongation rate is constant across growth conditions distinguishes *S. cerevisiae* from the bacteria *E. coli*, where the peptide elongation rate increases with increasing growth rate from 8 to 16 aa/s [12,16].

### tRNA is in excess in fast growth condition

To explore why the peptide elongation rate is constant across growth conditions, we examined how translation machinery components vary with the growth rate. We first quantified tRNA concentrations under five conditions by RT-qPCR, relative to three housekeeping genes with stable mRNA concentration across all five tested conditions [46]. Almost all nuclear-encoded tRNA increases concentration with increasing growth rate and is proportional to the concentration of active ribosomes (**Fig. 3g**; **Extended Data Fig. 4a-b**). Translation-related proteins all increase in concentration with increasing growth rate, but the elongation factors eEF1A, eEF1B, eEF2, and several initiation factors may increase in concentration more slowly than eEF3 and tRNA synthetases (**Fig. 3h; Extended Data Fig. 4c**).

That the peptide elongation rate remains constant suggests that charged tRNA are in excess and not limiting for growth. In other words, reducing the concentration of charged tRNA complexes should initially not impact cell growth. We sought to test this hypothesis by modulating the concentration of available charged tRNA. To do this, we manipulated eEF1A expression by deleting *TEF2* and replacing the promoter of *TEF1* with an aTc-inducible $P_{tet}$ promoter [47] (**Fig. 3i**). Reduction of available eEF1A reduces the export of tRNA from the nucleus, and thereby reduces the available charged tRNA for elongating ribosomes [48–50]. We expect that this manipulation should reduce the concentration of charged tRNA-eEF1A complexes in the cytoplasm because in glucose ~90% of the tRNAs are charged [51]. Remarkably, reducing eEF1A levels to 30% of the original level did not affect growth rate, the fraction of active ribosomes, or the peptide elongation rate (**Fig. 3j; Extended Data Fig. 4d-e**). Further reduction of eEF1A is lethal and we were unable to measure any intermediate states. Taken together, these experiments suggest that tRNA-complex availability is in excess in nutrient-rich environments so that even when they are reduced in slower growth conditions, the peptide elongation rates remain unaffected.

### Global mRNA concentration increases with growth rate

While most previous experimental studies of cell growth have focused on ribosomes as being limiting, theoretical work shows it is also possible for mRNA to be limiting [52,22]. Indeed, proteomic analysis revealed that the proteins exhibiting the most significant differences across growth conditions were primarily associated with both ribosomes and mRNA synthesis (**Extended Data Fig. 5a**). To test if growth rates drove changes in mRNA concentration, we performed spike-in RNA sequencing on yeast grown in four carbon sources with distinct growth rates. This showed that the total mRNA concentration increases monotonically with the growth rate (**Fig. 4a; Extended Data Fig. 5b-e**).

In contrast, global protein concentrations were similar across the 4 conditions (**Extended Data Fig. 5f**). For each gene, we observed that the protein fraction was close to proportional to the corresponding mRNA fraction across the conditions even though the global mRNA concentration changes (**Fig. 4b; Extended Data Fig. 5g**). This suggests a picture in which a global protein concentration is maintained and that the relative concentrations of specific proteins is largely determined by their relative mRNA concentrations [53]. However, the coupling between the fraction of protein and mRNA becomes weaker at lower growth rates ($r$ = 0.82 in glucose and $r$ = 0.62 in trehalose, **Extended Data Fig. 5g**), suggesting that protein expression may not be solely determined by transcription under slow-growth conditions. Taken together, our data suggest that global mRNA concentrations increase with growth rate in *S. cerevisiae* similar to what was found for bacteria [12].

### Mass action kinetics of RNAP II and the genome may explain growth-dependent increases in mRNA concentrations

To explore why mRNA concentration increases with growth rate, we first examined the concentrations of RNA polymerase II (RNAP II) subunits and their corresponding mRNA and found that they all slightly increased with the growth rate (**Fig. 4c and Extended Data Fig. 6a-b**). Since mRNA concentrations result from the balance of synthesis and degradation, we sought to determine whether mRNA decay rates varied across carbon sources. To do this, we acutely repressed the transcription of the methionine regulated genes *MET3* and *MET17* and measured their mRNA concentrations by RT-qPCR in all conditions (see **Methods**). mRNA decay rates of *MET3* and *MET17* were nearly the same in all tested conditions (**Fig. 4d and Extended Data Fig. 6c**). Next, to obtain a transcriptome-wide measurement of mRNA decay rates, we employed SLAM-seq for yeast using 4tU [54], a metabolic RNA labeling method. This allows us to determine the mRNA decay rates of each gene in fast-growth glucose and slow-growth trehalose [55] (**Fig. 4e and Extended Data Fig. 6d**, see **Methods**). Although individual mRNA decay rates may vary, the average mRNA decay rate shows only a slight increase in glucose relative to trehalose, despite a significant increase in growth rate (**Fig. 4f-g**). Therefore, we conclude that the increase in mRNA concentration with growth rate is primarily driven by increases in transcription rather than degradation.

However, the increase in RNAPII concentration alone cannot explain the 4.5-fold increase in total mRNA concentration in cells growing on glucose compared to trehalose as a carbon source. To better understand how increasing RNAP II concentrations drive increasing mRNA concentrations, we built a mathematical model (**Fig. 4h**). In our model, the concentration of DNA-bound RNAP II, $[pol_{bound}]$, is determined by mass action kinetics of free RNAP II, $[pol_{free}]$, and the nuclear genome concentration, $[DNA]$. Dissociation of the DNA-bound RNAP II is modeled using first order kinetics with off-rate $k_{off}^{pol}$ so that:

$$\frac{\mathrm{d}[pol_{bound}]}{\mathrm{d}t} = k_{on}^{pol}[DNA][pol_{free}] - k_{off}^{pol}[pol_{bound}] \text{ and}$$

$$[pol_{total}] = [pol_{free}] + [pol_{bound}] = \frac{P\phi_{pol}}{f},$$

where $P$ is the global protein concentration, $\phi_{pol}$ is the RNAP II fraction in proteome and $f = V_{nucleus}/V_{cell}$ denotes the ratio of nuclear volumes $V_{nucleus}$ to cell volumes $V_{cell}$. At steady-state, these two equations can be solved for the fraction of RNAP II that is DNA-bound:

$$\frac{[pol_{bound}]}{[pol_{total}]} = \frac{1}{1+\frac{K_d^{pol}}{[DNA]}} = \frac{1}{1+\frac{K_d^{pol} V_{cell} f}{DNA}},$$

where $K_d^{pol} = k_{off}^{pol} / k_{on}^{pol}$ is the dissociation constant of RNAP II, and *DNA* is the total amount of genome per cell. mRNA is produced by bound RNAP II in the nucleus and then exported to the cytoplasm so that

$$\frac{\mathrm{d}[mRNA]}{\mathrm{d}t} = k_{trans}^{mRNA}[pol_{bound}]f - k_d^{mRNA}[mRNA],$$

where $k_{trans}^{mRNA}$ is the mRNA transcription rate and $k_d^{mRNA}$ is the average degradation rate. At steady state, we can solve these equations for the mRNA concentration to yield:

$$[mRNA] = \frac{k_{trans}^{mRNA}}{k_d^{mRNA}}[pol_{total}]\phi_{pol}^{bound} f = \frac{k_{trans}^{mRNA} P \phi_{pol}}{k_d^{mRNA}\left(1+\frac{K_d^{pol} V_{cell} f}{DNA}\right)}.$$

We independently measured the global protein concentration, RNAP II fraction in proteome, DNA content, cell and nuclear volumes, and mRNA decay rate for 4 growth conditions. To measure the nuclear volume fraction $f$ for each condition, we used Pus1-eGFP as a nuclear marker. $f$ changes significantly depending on the growth condition, which impacts the nuclear concentrations of the genome and RNAP II (**Fig. 4i-j and Extended Data Fig. 6e-f**). We further experimentally determined the fraction of DNA-bound RNAPII using single-molecule tracking of the largest subunit, *RPB1-Halo* using methods we previously established [25]. DNA-bound RNAP II shows a slow diffusion coefficient, while free RNAPII diffuses rapidly. In glucose, approximately 55% of RNAPII is bound to DNA, consistent with previous reports [25,56]. In slow growth conditions, the fraction of DNA-bound RNAPII decreases to 30% (**Fig. 4k and Extended Data Fig. 6g**).

Our measurements greatly constrain our model so that we are left with only one free parameter, $K_d^{pol}$, to fit the fraction of RNAP II that is DNA bound. mRNA concentrations are then scaled with a second free parameter, the mRNA transcription rate $k_{trans}^{mRNA}$. Our two fit parameters are the same for all 4 growth conditions. Our simple transcription model recapitulates the observed increase in the fraction of DNA-bound RNAPII with growth rate and increasing mRNA concentrations (**Fig. 4l-m**). These results indicate that the increase in total mRNA concentration under fast growth condition is driven by increases in both RNAPII concentration and the DNA-bound fraction of RNAPII, with higher DNA concentration contributing to the increased proportion of RNAPII bound to DNA.

Our simple model assumes that transcription dynamics are the same in all these growth conditions and that the total amount of transcription is proportional to the amount of DNA-bound RNAPII, which is due to mass action kinetics of free nuclear RNAPII and the genome. Although we did not explicitly measure the elongation rate of RNAPII, our data are consistent with the rate remaining unchanged across the growth conditions examined. This model is similar to the one we previously used to explain how the total amount of transcripts scaled in proportion to the cell volume in a given growth condition [25].

## Mass action kinetics of free ribosomes and mRNA determine cellular growth

Since both mRNA and ribosome concentrations increase with growth rate, we hypothesize that mRNA-ribosome association and initiation kinetics may determine the fraction of active ribosomes and thereby determine the cellular growth rate. To test this hypothesis, we developed a dynamic equilibrium model describing mRNA-ribosome association and dissociation with mass action kinetics and first order decay, respectively (**Fig. 5a**). During translation initiation, the small ribosomal subunit binds near the 5′-end of the mRNA, scans to the start codon, and recruits the large subunit to form the complete 80S ribosome [57]. For simplicity, our model treats the 80S ribosome as a single unit that binds to an mRNA binding site, undergoes translation initiation, and then elongation. The ribosome then dissociates upon completion of translation and recycles back into the free pool. Assuming that elongating ribosomes do not saturate the mRNA, the concentration of ribosomes sitting in the initiating position, $\left[ribo_{init}\right]$, is determined by the concentration of free ribosomes, $\left[ribo_{free}\right]$, the free ribosome binding sites on the mRNA, $\left[RBS_{free}\right]$, the associated binding rate, $k_{on}^{ribo}$, and the rate of initiating translation, $k_{init}^{ribo}$, so that:

$$\frac{\mathrm{d}\left[ribo_{init}\right]}{\mathrm{d}t} = k_{on}^{ribo}\left[RBS_{free}\right]\left[ribo_{free}\right] - k_{init}^{ribo}\left[ribo_{init}\right].$$

After initiation, ribosomes form part of a translating pool whose concentration $\left[ribo_{trans}\right]$ is determined by the initiation rate and the termination rate, $k_{off}^{ribo}$, so that:

$$\frac{\mathrm{d}\left[ribo_{trans}\right]}{\mathrm{d}t} = k_{init}^{ribo}\left[ribo_{init}\right] - k_{off}^{ribo}\left[ribo_{trans}\right].$$

In addition, we can write the total ribosome concentration, $\left[ribo_{total}\right]$, as the sum of the distinct pools of ribosomes so that

$$\left[ribo_{total}\right]=\left[ribo_{free}\right]+\left[ribo_{init}\right]+\left[ribo_{trans}\right].$$

Finally, we use the fact that the mRNA concentration equals to concentration of free ribosome binding sites (only one per mRNA) and concentration of occupied ribosome binding sites:

$$\left[mRNA_{total}\right]=\left[ribo_{init}\right]+\left[RBS_{free}\right].$$

We then scale our variables to make the equations dimensionless and solve them for steady state solutions (see **Methods** and **Extended Data Fig. 7a**). This results in a set of equations with a single free parameter, $\alpha_{glc}$, which is the dimensionless mRNA concentration in glucose. Using our single free parameter, we fit our data for the active ribosome fraction as a function of the cellular growth rate in all our conditions (**Fig. 5b; Extended Data Fig. 7b-c**). Taken together, our analysis shows that the fraction of active ribosomes could be determined by simple mass action kinetics (**Fig. 5c**).

To directly test our model that the fraction of active ribosomes is determined by mRNA and ribosome mass action kinetics, we sought to reduce the mRNA concentration. To do this, we again used the anhydrotetracycline (aTc)-inducible promoter [47] to control the expression of the RNAP II subunit *RPB1* so that it ranges from negligible amounts to above the wild type level (**Fig. 5d**). This yeast strain cannot survive without aTc, and its growth rate increases with aTc dosage, reaching a maximum at approximately 30 ng/mL (**Extended Data Fig. 7d**). Spike-in mRNA sequencing showed that global mRNA concentrations rose from 0.4-fold to near-normal levels as aTc dosage increased (**Fig. 5e**). Lowering mRNA concentrations reduced the fraction of active ribosomes as predicted by our model (**Fig. 5f-g**).

### Yeast cells coordinate mRNA and ribosomes to regulate growth rate

Our model is easily extended to predict the growth rate, $\lambda$, which is proportional to the concentration of active ribosomes since the peptide elongation rate is constant across different conditions:

$$\lambda=\frac{c_p\phi_a\phi_r}{m_{rb}}.$$

Thus, for the four conditions where we have measured all the model's parameters, we agreement with our experimental results (**Fig. 6a**). In principle, the growth rate can be increased by either increasing any combination of the mRNA or ribosome concentrations. However, *S. cerevisiae* appears to coordinate this regulation so that the ribosome to mRNA ratio is constant (**Fig. 6b**).

We further tested our growth model by examining a naturally occurring situation where the mRNA concentration decreases, but the ribosome concentration remains constant. This occurs when yeast cells are arrested in the G1 phase of the cell cycle for extended periods of time [25,58] (**Extended Data Fig. 7e**). In this case, we accurately predict how the growth rate declines with the decreasing mRNA concentration (**Fig. 6c; Extended Data Fig. 7f-g;** see **Methods**).

Perhaps the most surprising prediction of our mRNA–ribosome kinetic model is that growth rate would be increased by increasing the mRNA concentration. We therefore sought to test this prediction by conditionally inhibiting mRNA degradation to increase mRNA concentration. To do this, we fused an AID (auxin-inducible degradation) tag to the mRNA decapping enzyme Dcp2, which allowed us to trigger its conditional degradation by adding auxin to the media [59] (**Fig. 6d; Extended Data Fig. 7h**). Following auxin addition, we measured the total mRNA concentration at minute resolution with fluorescence in situ hybridization (FISH) using a 30-mer poly-dT (T30) probe (see **Methods**). Total mRNA concentration increased within the first 20 minutes after auxin addition (**Fig. 6e**). Moreover, this increase in mRNA concentration transiently enhanced the cellular growth rate in line with the model prediction based on the assumption that ribosome content remains constant over short timescales (**Fig. 6f**; **Extended Data Fig. 7i**). After a transient increase, the growth rate slows, likely due to the toxic effects of Dcp2 degradation that are expected because of the deleterious effects of *dcp2*Δ [60].

## DISCUSSION

Our measurements reveal that budding yeast modulates growth by coordinately increasing both ribosome and mRNA concentrations, thereby raising the fraction and absolute number of active ribosomes that translate at an approximately constant elongation speed. These results highlight a conserved principle across bacteria and yeast: growth is strongly influenced by the abundance of actively translating ribosomes, which depends on the availability of both ribosomes and mRNA [12,28]. In the conditions we examined, changes in yeast growth rate are explained primarily by the proportional scaling of ribosome and mRNA concentrations, whereas average elongation speed remains nearly constant. This differs from *E. coli*, where elongation speed increases with growth rate [14]. While changes in elongation rate appear to make a more minor contribution to overall growth, this difference suggests that the underlying regulatory architectures governing growth may differ between yeast and bacteria. These findings establish a new quantitative framework for eukaryotic growth in which it is determined both by global transcriptional output and ribosome biogenesis.

Mechanistically, our framework rests on two simple mass action kinetic models (**Fig. 6g**). In the first mass action model, RNAP II concentration increases with the growth rate to increase its rate of association with the genome. This drives a proportional increase in the concentration of total mRNA. For the total mRNA concentration to be a function of the cell growth rate in different conditions, it must evade the general homeostatic mechanisms that act to ensure mRNA concentration homeostasis. For example, feedback mechanisms ensure that mRNA degradation decreases in response to mutations lowering the transcription rate so that the mRNA concentration stays the same [23,25,61,62]. In contrast, we find that mRNA half-lives are not modulated in response to different growth conditions so the global mRNA concentration changes (**Fig. 4g**). In the second mass action model, ribosome biogenesis increases in proportion to the growth rate so that the active fraction of ribosomes is determined by mass action kinetics with mRNA. As mRNA and ribosome pools rise in lock-step, the concentration of active ribosomes increases to determine the growth rate.

These results have unexpected evolutionary implications. First, since charged-tRNA are buffered, it should be possible to increase the growth rate by increasing either the ribosome concentration or the mRNA concentration, even in rich nutrient conditions such as glucose containing media. Indeed, even our conditional increase in mRNA concentration led to a transient increase in the growth rate. That yeast does not do so, after decades of propagation in the laboratory, suggests that laboratory conditions represent a more complex ecology than is generally thought. Consistent with this notion, evolution experiments are quickly able to select for more fit mutants in specific environments [63,64]. Second, to increase growth rate, yeast coordinately upregulate both ribosome and mRNA concentrations. It is not obvious why it does so because the most efficient way would be to simply increase mRNA, which is 1-2% of the cell's dry mass, rather than ribosomes, which are over 10%. Thus, while it is possible to produce a particular growth rate with a continuous range of ribosome and mRNA concentrations, yeast appear to be constrained to maintain about 2 ribosomes per mRNA.

While the mechanistic reason for the coordinated regulation of mRNA and ribosomes can be understood as being due to both ribosome biogenesis and RNAP II synthesis being downstream of a common mTOR activity [65,66], the evolutionary reasons for this apparent constraint are currently a mystery. One possibility is that the regulation of eukaryotic growth will be better understood when we consider dynamic environments requiring rapid adaptation. While mRNA concentrations can be rapidly tuned on a 10-minute timescale, ribosome concentrations can only slowly be tuned on a timescale of hours. It could be possible that maintaining a higher concentration of ribosomes better prepares the cell to take advantage of rapid upshifts in nutrients, while making growth also dependent on the quickly turned over mRNA allows the cell to quickly respond to worsening conditions. Certainly, more work needs to be done to understand how growth, perhaps the most fundamental physiological property of a cell, is coordinated in response to environmental conditions. While our work is an important step forward in this path, we anticipate future work extending our single-molecule tracking and spike-in transcriptomic methods to other eukaryotic cells to test the generality of our framework and refine models of cellular growth.

## Acknowledgements

ZY478 and ZY483 are gifts from the Zid lab. This work was supported by the National Institute of Health R35 GM134858. We thank Huan Zheng, Skye Glenn, and Cristhian Chumpitaz for teaching us the single molecule methods used in this work and Christine Jacobs-Wagner and for the generous use of their STORM microscope. We thank Magda Wąchalska for help in polysome profile. We thank Jennifer Ewald for stimulating discussions about cell growth and metabolism.

## Declaration of interests

The authors declare no competing interests.

## Data and code availability

Original code (i.e., the modelling and simulations) has been deposited at Github: https://github.com/gaox0826/mRNA-ribosome-model.git. Raw sequencing data (RNA-seq, ribosome footprinting, and SLAMseq) are publicly available in the NCBI Sequence Read Archive (SRA) under BioProject PRJNA1359129. Mass-spectrometry proteomics data have been deposited with the ProteomeXchange Consortium via PRIDE (Project accession: PXD070570). Processed transcriptomic and proteomic datasets used to generate the figures are provided as Supplementary Data tables. Additional information and materials are available from the corresponding author upon reasonable request.

## Methods

### Yeast strains and media conditions

All *S. cerevisiae* strains used in this study are in the *W303 MATa* background. The genotypes of all strains are listed in **Supplementary Table 1**. XG03, XG04, XG06 and XG07 were used to measure the active ribosome fraction by single molecule tracking (SMT). A sucrose cushion analysis confirmed that the Halo tag was successfully incorporated into the 80S ribosome. Strain XG48 ($P_{tet}$-*RPB1*) was used to assess changes in the active ribosome fraction under reduced mRNA levels. In this strain, the repressors TetR and TetR-Tup1 were integrated into the *URA3* locus, and an inducible promoter ($P_{7tet.1}$) was placed upstream of the *RPB1* gene [47]. Similarly, strain XG61 ($P_{tet}$-*TEF1*, *tef2Δ) was* used to change eEF1A expression. MS165 was used to measure the DNA-bound Rpb1 fraction by SMT [25]. Strain ML127 was used for the G1 arrest experiments [58]. ZY478 and ZY483 were used to measure the onset of luminescence from Nluc reporters with variable-length upstream coding sequences under different carbon *sources* [44,45]*.* XG72 integrates a *TIR1–Myc* allele at the *URA3* locus and fuses *AID–Myc* to the C-terminus of *DCP2*, enabling auxin-dependent degradation of Dcp2.

For carbon-limited conditions, yeast cells were cultured in medium prepared with low-fluorescence yeast nitrogen base (LF-YNB, Formedium CYN6205) to reduce background fluorescence in single-molecule experiments, along with a complete amino acid dropout mix. One of the following 12 carbon sources (2% w/v) was added as the sole carbon source: fructose, glucose, sucrose, mannose, acetate, galactose, lactate, ethanol, pyruvate, glycerol, trehalose, or melibiose. For nitrogen-limited conditions, yeast cells were cultured in a medium prepared with LF-YNB and supplemented only with the minimal required auxotrophic supplements: tryptophan, uracil, and histidine. One of the following carbon and nitrogen source combinations were added: 2% glucose + 2 mM glutamic acid, 2% galactose + 2 mM glutamic acid, or 2% glucose + 2 mM leucine.

The pH of all media was adjusted to 6.0 using NaOH. For G1 arrest experiments, cells were grown in SC + 2% glycerol+ 1% ethanol. For the XG48 ($P_{tet}$-*RPB1*) strain, cells were grown in SC + 2% glucose and 5-200 ng/ml aTc. Unless stated otherwise, cells were grown at 30 °C for at least 10 doublings before the experiment was conducted. Cells were harvested at an OD600 between 0.1 and 0.4 for subsequent measurements.

### Growth rate measurements

To measure the growth rate of cells in a specific growth condition, cells were cultured to the exponential phase and then diluted into a pre-heated flask with 50 ml of fresh media (OD600=0.01). Cells were incubated at 30 °C and OD600 was measured every 1-2 hours. To determine the specific growth rate, log-transform OD600 values were plotted against time, and the slope of this log(OD600) versus time plot was calculated as the specific growth rate. The population doubling time can then be calculated as ln2/growth rate. There were at least two replicates of the growth rate measurement for each condition. For the local growth rate measurement in the G1 arrest experiment, please see the detailed description in the “G1 arrest experiment” section.

**LC-MS/MS sample preparation and data acquisition**

A 10 mL yeast culture (OD600=0.2) was harvested, snap-frozen, and stored at -80 °C. 1mM PMSF and protease inhibitors (1 tablet/10 mL) were then added to the lysis buffer (50 mM Tris-HCl pH 8.0, 0.2% Tergitol, 150 mM NaCl, 5 mM EDTA). Cells were then resuspended from in the lysis buffer, and lysed using a FastPrep-24 (4 °C, 5.5 m/s, 35 s). The lysate was then centrifuged (15,000 × g, 5 min, 4 °C) and the supernatant collected. SDS was then added to 1%, DTT to 5 mM, and the sample was incubated at 65 °C for 15 min. Then, 15 mM Iodoacetamide was added before incubation at room temperature for 15 min. Proteins were then precipitated by adding >3 volumes PPT solution (acetone:ethanol:acetic acid = 50%:49.9%:0.1%) folloed by incubation on ice for 15 min and centrifugation (17,000 × g, 5 min, 4 °C). The supernatant was discarded and the pellet washed with 500 µL PPT solution. The sample was then centrifuged again, the supernatant discarded, and the pellet was air dried for ~10 min. The pellet was then resuspended in fresh 75 µL urea/Tris solution (8M urea, 50 mM Tris-HCl pH=8.0) before adding 225 µL 150 mM NaCl solution. The sample was digested overnight at 37 °C with trypsin (1:50 enzyme:protein), with agitation. The next day, 15 µL 10% TFA (Trifluoroacetic acid) was added to stop digestion and the sample was centrifuged (17,000 × g, 2 min), and the supernatant recovered. The supernatant was then desalted using a C18 Cartridge (Waters #WAT054955). Aliquots were then stored at -20 °C.

Desalted peptides were analyzed on a Fusion Lumos mass spectrometer (Thermo Fisher Scientific) equipped with a Thermo EASY-nLC 1200 LC system (Thermo Fisher Scientific). Peptides were separated by capillary reverse-phase chromatography on a 25-cm column (75-µm inner diameter, packed with 1.6-µm C18 resin, cat. no. AUR2-25075C18A, Ionopticks). Peptides were introduced into the Fusion Lumos mass spectrometer using a 125-min stepped linear gradient at a flow rate of 300 nl $min^{-1}$. The steps of the gradient are as follows: 3–27% buffer B (0.1% (v:v) formic acid in 80% acetonitrile) for 105 min, 27–40% buffer B for 15 min, 40–95% buffer B for 5 min and finally maintaining at 90% buffer B for 5 min. Column temperature was maintained at 50 °C throughout the procedure. Xcalibur software (v.4.4.16.14, Thermo Fisher Scientific) was used for the data acquisition and the instrument was operated in data-dependent mode. Advanced peak detection was enabled. Survey scans were acquired in the Orbitrap mass analyzer (Profile mode) over the range 375–1,500 m/z with a mass resolution of 240,000 (at m/z 200). For MS1, the normalized AGC target (%) was set at 250 and maximum injection time was set to 'Auto'. Selected ions were fragmented by higher-energy collisional dissociation with normalized collision energies set to 31 and the fragmentation mass spectra were acquired in the ion trap mass analyzer with the scan rate set to 'Turbo'. The isolation window was set to the 0.7-m/z window. For MS2, the normalized AGC target (%) was set to 'Standard' and maximum injection time to 'Auto'. Repeated sequencing of peptides was kept to a minimum by dynamic exclusion of the sequenced peptides for 30 s. Maximum duty cycle length was set to 1 s.

All raw files were searched using the Andromeda engine embedded in MaxQuant (v.2.4.2). Variable modifications included oxidation (M) and protein amino-terminal acetylation. Carbamidomethyl (C) was a fixed modification. The number of modifications per peptide was capped at five. Digestion was set to tryptic (proline blocked). MaxLFQ was used to determine the summed ion intensity for each protein. The database search was conducted using the UniProt proteome—Yeast_UP000002311_559292. The minimum peptide length was seven amino acids. The 1% false recovery rate (FDR) was determined using a reverse decoy proteome.

### Measurement of total RNA concentration

In addition to quantifying the ribosomal fraction of the proteome, we measured how total RNA concentration changes with growth rate. For each condition, two identical aliquots of 5 mL exponentially growing yeast culture (OD600 = 0.3–0.5) were harvested by centrifugation, snap-frozen in liquid nitrogen, and stored at −80 °C. One aliquot was used for total RNA extraction, while the other aliquot was used for protein extraction. One aliquot was resuspended in 500 µL TRI Reagent and lysed using a FastPrep-24 instrument (5.5 m/s for 40 s at 4 °C), repeated twice. Lysates were clarified by centrifugation at 20,000 × g for 5 min at 4 °C, and the supernatant was retained. Then, 100 µL chloroform was added, mixed thoroughly, and incubated for 2 min at room temperature, followed by centrifugation at 12,000 × g for 15 min at 4 °C. Two hundred microliters of the aqueous phase containing RNA was recovered. RNA was precipitated with isopropanol, washed with 75% ethanol, and resuspended in 40 µL RNase-free water. RNA concentration was measured using a NanoDrop spectrophotometer.

To normalize total RNA concentration and enable comparison across different growth conditions, protein concentration was measured from a parallel sample derived from the same amount of starting culture. Briefly, culture was harvested, snap-frozen, and stored at −80 °C. Cell pellets were resuspended in urea lysis buffer and lysed using a FastPrep-24 instrument (5.5 m/s for 40 s at 4 °C). Lysates were clarified by centrifugation at 17,000 × g for 10 min at 4 °C, and the supernatant was collected. Protein concentration was determined using the Protein Assay Dye (Bio-Rad).

Normalized total RNA concentration was calculated as the ratio of total RNA concentration to total protein concentration measured from the same cell mass. Two biological replicates were measured for each condition, with two technical replicates per biological replicate.

**Immunoblotting**

Ten µl cultures were grown to an OD600 of 0.1, harvested, snap-frozen, and stored at –80 °C. Cell pellets were resuspended in urea lysis buffer and lysed using a FastPrep-24 instrument (5.5 m/s for 40 s at 4 °C). Lysates were clarified by centrifugation at 17,000 × g for 10 min at 4 °C, and the supernatant was recovered. Protein concentrations were determined using the Protein Assay Dye (Bio-Rad). Then, equal amounts of protein were loaded on 10% Tris-glycine SDS-PAGE gels and transferred to nitrocellulose membranes using the iBlot 2 Dry Blotting System (Invitrogen). The following primary antibodies were used for Immunoblotting at 1:1000 dilution: Anti-HaloTag (rabbit, Polyclonal, Promega #G928A), Anti-Rpb1 (mouse, monoclonal, Sigma-Aldrich #05-952-I), Anti-eEF1A (mouse, monoclonal, Sigma-Aldrich #05235), Anti-c-Myc (mouse, monoclonal, Sigma-Aldrich #M5546), and Anti-Actin (rabbit, polyclonal, Sigma-Aldrich #A2103). Primary antibodies were detected using the following fluorescently labeled secondary antibodies at 1/10,000 dilution: Alexa Fluor 680 Goat anti-Mouse (Invitrogen #A21058), IRDye 680RD Goat anti-Rabbit (LI-COR # 926-68071), Alexa Fluor 790 Goat anti-Rabbit (Invitrogen #A11369), and IRDye 800CW Goat anti-Mouse (LI-COR # 926-32210). Membranes were imaged on a Typhoon biomolecular imager.

**Single-molecule tracking to determine the active ribosome fraction and the DNA-bound RNAP II fraction**

To prepare samples for tracking single ribosome molecules, cells are cultured for more than 10 doublings before harvesting (OD600=0.1-0.3). JF-PA549 Halotag (Janelia Farms photoactivatable 549) dye was added at a final concentration of 1 nM for ribosomal protein-Halo fusion strains and 10 nM for *RPB1-Halo* strain [67]. Cultures were incubated at 30 °C for 40 min and then pelleted. Cells then washed three times in fresh media to remove unbound dye and resuspend in 15 µl media. A 65 µl gene frame (1.5*1.6 cm) was glued to the slide and the SC culture solution containing the specific carbon source was added. 4 µl of cell culture was placed on the surface of an agarose slide containing the nutrient condition of interest and spread evenly by a pipette tip or clean stick.

Live cell photoactivated localization microscopy (PALM) was performed on a Nikon N-STORM microscope equipped with a Perfect Focus System and a motorized stage. JF549 was detected using an iXon3 DU 897 EMCCD camera (Andor) and excited using a 50 mW 561 nm laser (MLC400B laser unit, Agilent) at 30 °C. The laser was focused through a 100x Apo TIRF 1.49 NA oil immersion objective (Nikon) onto the sample using an angle for highly inclined thin illumination to reduce background fluorescence. Fluorescence emission was filtered by a C-N Storm 405/488/561/647 laser quad set. Transmission illumination was used to gather brightfield images.

PALM experiments were performed by using continuous activation of molecules with low power (0.1% – 5% in NIS-element software) 405 nm light to photoactivate around 1-2 molecules/cell at the same time. Simultaneous fast-exposure (15 ms) illumination with 561 nm light (50% in NIS-element software) was then used to image molecules. PALM movies of 10,000 frames were acquired with continuous laser illumination and a camera frame rate of 37 ms. A bright-field picture was taken to determine the cell area.

After imaging, single particle tracking was performed by Trackmate on ImageJ software. Molecules were localized in each frame using a Laplacian of Gaussian (LoG) method with an estimated diameter of 1 µm. A trajectory allows a maximum of 2 µm of movement within a frame, and a single frame disappearance within a trajectory was allowed. The diffusion coefficient of a single molecule ($D_a$) can be calculated from the stepwise mean-squared displacement (MSD) of its trajectory:

$$D_a = \frac{1}{4n\Delta t}\sum_{i=1}^{n} [x(i\Delta t) - x(i\Delta t + \Delta t)]^2 + [y(i\Delta t) - y(i\Delta t + \Delta t)]^2$$

where x(t) and y(t) denote the coordinates of the ribosome of interest at time t. Δt is the time between two consecutive frames, and n is the total steps in the trajectory. The ribosome trajectories with less than 9 displacements were omitted.

**Polysome profile measurements**

Yeast cells were grown in 100 mL SC medium with different carbon sources at 30°C until the culture reached an OD600 of approximately 0.5. To stabilize ribosomes, cycloheximide (CHX) was added at a final concentration of 100 µg/mL for 5 min before harvesting. Cells were collected by centrifugation at 3,000×g for 5 min at 4°C, and the pellet was then stored at -80°C. The cell pellet was resuspended in a pre-chilled lysis buffer A (20 mM Tris-HCL, pH 7.4; 50 mM KCl; 10 mM $MgCl_2$; 1 mM DTT added immediately prior to use). Cells were lysed using a Fastprep 24 (settings: 5.5 m/s, 1 x 35 seconds) at 4 °C, followed by 5 min on ice. This process was repeated 6 times. The lysate was then clarified by centrifugation at 3,000×g for 5 min at 4°C, then by centrifugation at 11300×g, at 4°C for 2 and 10 min, respectively. The supernatant, which contains ribosomes, was then collected for gradient separation.

A 4.5–45% (w/v) sucrose gradient was prepared in buffer A in a 13.2 mL tube using a Biocomp Gradient Master. The sucrose gradient was stored at 4°C for at least 1 hour. Equivalent numbers of OD260 units (10 units) in a total volume of 200 ml Buffer A was then loaded on the 4.5–45% sucrose gradients. 200 ml Buffer A without cell lysis was also loaded on a sucrose gradient as a blank. Then, the samples were centrifuged at 35,000 rpm for 2 hours and 10 min at 4°C on a SW41 Ti rotor in a Beckman ultracentrifuge. After centrifugation, the gradient is fractionated from the top, and absorbance at 254 nm is continuously monitored to reveal the polysome profile by a Biocomp Piston Gradient Fractionator.

To estimate the active ribosome fraction by polysome profile data, we first subtracted the signal of a blank sample from the signal absorbance at 254 nm. For 40S and 60S ribosome subunits, the signal was integrated from the region indicated on **Extended Data Fig. 2c**. Because each ribosome contains roughly 5500 bp of rRNA and the average mRNA length is about 1250 bp, the rRNA fraction in the monosome is approximately 82% [4]. Thus, we multiplied the integrated monosome signal by 0.82 to obtain its rRNA contribution. For polysomes containing two ribosomes, a factor of 0.90 was used. Similarly, larger polysomes were corrected using their respective rRNA fractions. All polysomes were assumed to be active, and all other ribosomes were considered inactive to calculate the active ribosome fraction. Accordingly, the active ribosome fraction was calculated as polysome signal over total signal: (polysome)/(40s+60s+monosome+polysome).

### Estimate average peptide elongation rate of active ribosomes

We estimate the average peptide elongation rate of a single ribosome by the synthesis rate of the total protein. Since yeast grows exponentially, the protein mass increases at a rate proportional to the existing protein mass, $m_{prot}$, so that $\frac{dm}{dt} = \lambda m = \frac{m \ln 2}{T_d}$, where $\lambda$ is the growth rate and $T_d$ is the population doubling time. The rate of protein mass increase is given by the number of all ribosomes, $N_{ribo}$, and the average peptide elongation rate for all ribosomes, $e_r$, so that:

$$e_r = \frac{m_{prot} \ln 2}{N_{ribo} T_d}.$$

We now consider the fact that only some ribosomes are active. If the active ribosome fraction is $\phi_a$, and the average elongation rate of an active ribosome $c_p$ is:

$$c_p = \frac{m_{prot} \ln 2}{\phi_a N_{ribo} T_d}.$$

We define the ribosome fraction in the proteome as $\phi_r$, and the average protein mass of a single ribosome as $m_{rb}$ so that the average ribosomal protein mass in the cell population is:

$$m_{prot} \phi_r = N_{ribo} m_{rb}.$$

Then, the average elongation rate of an active ribosome can be rewritten as

$$c_p = \frac{m_{rb} \ln 2}{\phi_a \phi_r T_d}.$$

The average protein mass of a single ribosome $m_{rb}$ is ~11984 amino acids [14,68]. The remaining three parameters (ribosome fraction in the proteome $\phi_r$, active ribosome fraction $\phi_a$ and doubling time $T_d$) were obtained from the measurements in this study. Since $\frac{\ln 2}{\lambda} = T_d$, the growth rate $\lambda$ can be calculated as:

$$\lambda = \frac{c_p \phi_a \phi_r}{m_{rb}}.$$

### Ribosome footprinting

Yeast cells were cultured in 200 mL synthetic complete (SC) medium containing various carbon sources at 30 °C until the cultures reached an OD600 of ~0.5 for ribosome footprinting [69]. Cells were harvested by vacuum filtration (GenClone), scraped from the filter membrane with a spatula, immediately snap-frozen in liquid nitrogen, and stored at –80 °C. Cell pellets were resuspended in 300 µL of ice-cold lysis buffer (20 mM Tris-HCl, pH 7.4; 150 mM NaCl; 5 mM $MgCl_2$; 1 mM DTT; 1% [v/v] Triton X-100; 25 U/mL

Turbo DNase I; 100 ug/ml cycloheximide) and lysed using a FastPrep-24 instrument (5.5 m/s for 35 s) at 4 °C. After each 35 s cycle, samples were incubated on ice for 5 min; this cycle was repeated six times. Lysates were clarified by centrifugation at 3,000 × g for 5 min at 4 °C, and the supernatant was transferred to a fresh tube and centrifuged again at 20,000 × g for 10 min at 4 °C. Then, 800 µg of total RNA was diluted to 200 µL with lysis buffer and treated with 1.5 µL RNase I for 45 min at room temperature. Digestion was terminated by adding 10 µL RNase Inhibitor (SUPERase·In, Invitrogen). A 4.5–45% (w/v) sucrose gradient was prepared in polysome buffer (20 mM Tris-HCl, pH 7.4; 150 mM NaCl; 5 mM $MgCl_2$; 1 mM DTT) in a 13.2 mL ultracentrifuge tube using a Biocomp Gradient Master and equilibrated at 4 °C for at least 1 h.

Two hundred µL of the RNase-treated lysate was gently layered onto each gradient and centrifuged at 36,000 rpm for 2 h at 4 °C on a SW41 Ti rotor in a Beckman Optima XE90 ultracentrifuge. Gradients were then fractionated from the top while continuously monitoring A254 with a Biocomp Piston Gradient Fractionator. Then, the fraction containing monosomes was collected. Monosome fractions were adjusted to 1% (w/v) SDS and 200 µg/mL proteinase K (Thermo), mixed by gentle inversion, and incubated at 42 °C for 30 min. Phenol:chloroform:isoamyl alcohol (25:24:1) was added to the proteinase K–treated fractions in a 1:1 ratio, vortexed vigorously, and centrifuged at 16,000 × g for 5 min at room temperature. The upper aqueous phase was transferred to a fresh tube, and RNA was precipitated by adding 0.1 volume of 3 M sodium acetate (pH 5.2), 2.5 volumes of 100% ethanol, and 1 µL glycogen, followed by incubation at –20 °C overnight. Precipitated RNA was pelleted by centrifugation at 20,000 × g for 30 min at 4 °C. The supernatant was then carefully removed, and the pellet was air-dried on ice for 10 min. Finally, RNA was resuspended in 5 µL of 10 mM Tris (pH 8.0). Samples were mixed 1:1 with RNA loading dye (NEB B0363S), denatured at 80 °C for 90 s, and resolved on a 15% polyacrylamide TBE-urea gel at 200 V for 60 min. Gels were stained with 1× SYBR Gold in TBE running buffer, visualized under UV illumination, and the 17–34 nt region was excised into a fresh tube. Excised gel slices were incubated in 400 µL RNA extraction buffer (300 mM sodium acetate, pH 5.5; 1 mM EDTA; 0.25% [v/v] SDS) overnight at room temperature to elute the RNA. The eluate was precipitated with ethanol (or isopropanol) as described above, and then pelleted by centrifugation, air-dried on ice, and resuspended in 4 µL of 10 mM Tris-HCl (pH 8.0). Recovered RNA (3.5 µL) was treated with T4 polynucleotide kinase by adding 0.5 µL 10× T4 PNK buffer, 0.5 µL T4 PNK (NEB), and 0.5 µL SUPERase•In, then incubating at 37 °C for 30 min to repair 2′–3′ cyclic phosphates. NEBNext Small RNA Library Prep Set (E7330S) was used to build libraries for pair-end (2×150 bp) Illumina sequencing (GENEWIZ). 15 million reads were recorded per sample.

Raw paired-end reads were first processed with Trimmomatic v0.39 to remove adapter sequences and perform quality trimming. We retained only reads ≥ 15 nt. Low-quality bases were trimmed using default parameters. Trimmed reads were then aligned against the *S. cerevisiae* rRNA reference sequence (SGD S288C; http://sgd-archive.yeastgenome.org/sequence/S288C_reference/rna/) using Bowtie v1.2.3 with options -f -v 2 -k 1 -p 4. Reads mapping to rRNA were discarded. The remaining non-rRNA reads were mapped to the *S. cerevisiae* genome (NCBI R64) with Bowtie (-m 1) to retain only uniquely aligned fragments. Finally, these uniquely mapped reads were aligned to the *S. cerevisiae* transcriptome (NCBI R64) to produce BAM files for subsequent analysis of ribosome footprints to calculate translation efficiency.

**Calculation of translation efficiency**

Here, we calculate the translation efficiency of each gene based on spike-in RNA-seq and ribo-seq across four growth conditions [43]. The normalized ribosome footprint density for each gene was obtained by dividing the number of ribosome footprint reads mapped to its coding sequence (CDS) by the CDS length, and then normalizing this value to the total number of ribosome footprint reads mapped to the yeast transcriptome. Similarly, the normalized mRNA read density was calculated by dividing the number of mRNA reads mapped to the CDS by the CDS length, followed by normalization to the total mRNA reads. Genes with fewer than 20 reads were excluded from the analysis. The translation efficiency (TE) of each gene in each condition can then be calculated as:

$$TE = \frac{norm.\ RFs\ density}{norm.\ mRNA\ density}.$$

**Measure the translation onset time using a luciferase reporter**

Two constructs were analyzed: $P_{tet}$–*Nluc* (ZY478) and $P_{tet}$–*YFP–Nluc* (ZY483), the latter containing a 315-amino-acid YFP insert upstream of Nluc [44]. Cells were grown in SC medium containing either glucose or glycerol at 30 °C to an OD600 of 0.1. For each strain and condition, 150 µL of culture was dispensed into wells of a black 96 well plate (Cellvis). One set of wells was supplemented with 250 ng/mL anhydrotetracycline (aTc) and 5 µM furimazine, while control wells received only 5 µM furimazine. Plates were incubated in a microplate reader (BioTek) at 30 °C, and bioluminescence along with OD600 was measured every 30 s for 60 min, with a 3 s orbital shake prior to each reading. Three biological replicates were performed for each condition.

Raw bioluminescence values were first normalized to cell density by dividing each reading by the corresponding OD600. Next, luminescence in the presence of aTc was background-corrected by subtracting the signal from the uninduced control. These background-corrected values were then baseline-corrected using the mean of the first five time points. Relative light units (RLU) were then square-root transformed. Translation onset time was estimated by performing linear regression on the square-root transformed data in its linear region (Schleif plot). The x-intercept of this fit approximates the time required to translate the upstream coding sequence.

### RT-qPCR to determine the tRNA abundance

We measured tRNA abundance by RT-qPCR in yeast cells cultured with five different carbon sources (glucose, mannose, galactose, glycerol, and trehalose). Cultures were grown to an OD600 of 0.1, and 20 mL of each culture was harvested, snap-frozen, then stored at –80 °C. Cells were lysed at 4 °C using a FastPrep 24 instrument (5.5 m/s, 40 s), and total RNA was extracted with the Direct-zol RNA Microprep Kit (Zymo Research). All RNA samples had A260/A230 and A260/A280 ratios above 2.

For cDNA synthesis, 1 µg of RNA was reverse-transcribed using SuperScript VILO Master Mix (Thermo Fisher), with the reaction carried out at 60 °C for 30 min to mitigate potential effects of tRNA secondary structures and modifications. The resulting cDNA was stored at –20 °C until further use. One µL of cDNA was diluted 1:100 for qPCR, which was performed using iTaq Universal SYBR Green Supermix (Bio-Rad #1725121). The 42 tRNA primer pairs we used were taken from Torrent et al [46]. Two biological replicates were performed per experiment and 2 technical replicates were performed per biological replicate.

To normalize tRNA levels, we selected three housekeeping genes (*PLN1*, *RNR2*, and *PRC1*), whose mRNA concentrations remained consistent across the different carbon sources, as indicated by our spike-in mRNA sequencing data (**Extended Data Fig. 4b**). The geometric mean of these housekeeping genes served as the reference for calculating tRNA concentrations, and relative expression levels were determined using the standard $\Delta\Delta C_t$ method.

### Spike-in RNA-seq to determine total mRNA concentration

Global mRNA concentration in different growth conditions was quantified by spike-in RNA sequencing. Ten milliliters of *S. cerevisiae* cultures (OD600=0.1-0.2) in different growth conditions were harvested by centrifugation at 12,000 × g for 30 s, snap-frozen in liquid nitrogen, and stored at –80 °C. As an external spike-in control, 5 mL *C. glabrata* was grown in SC medium with 2% glucose to an OD600 of ~0.2, then harvested, snap-frozen, and stored at –80 °C.

For spike-in normalization, the *C. glabrata* pellet was resuspended in 50 µL ice-cold PBS and combined with the *S. cerevisiae* pellet by gentle mixing. Twenty microliters of the mixed cells were removed and stored at –20 °C for genomic DNA sequencing. The remaining 30 µL was lysed in 300 µL TRI Reagent (Zymo Research) for total RNA extraction, following the manufacturer's protocol.

For total RNA sequencing, 30 µL of mixed cells were lysed in a FastPrep-24 instrument (5.5 m/s for 35 s at 4 °C). Lysates were clarified by centrifugation at 20,000 × g for 5 min at 4 °C, and the supernatant was retained. Total RNA was purified using the Direct-zol RNA Microprep Kit (Zymo Research) following the manufacturer's instructions. Purified RNA was submitted to GENEWIZ for mRNA library preparation and sequenced on an Illumina HiSeq (2×150 bp), yielding >20 million paired-end reads per sample. Raw reads were trimmed with fastp v0.23.1 to remove adapter sequences and low-quality bases. Clean reads were quantified using Salmon (v1.10) against a combined transcriptome reference consisting of *Saccharomyces cerevisiae* (SGD) and *Candida glabrata* (CGD assemblies). Transcript abundance was estimated in transcripts per million (TPM) for downstream analysis.

Genomic DNA was extracted using the YeaStar Genomic DNA Kit (Zymo Research) according to the manufacturer's chloroform protocol. Purified gDNA libraries were prepared and sequenced by GENEWIZ on an Illumina HiSeq platform (2×150 bp), generating >5 million paired-end reads per sample. Paired-end reads were trimmed to remove adapters and low-quality bases using *Trimmomatic* (v0.39). A combined *Saccharomyces cerevisiae* (sacCer3, R64) and *Candida glabrata* (CGD assemblies) reference genome was built, and trimmed reads were aligned to this reference using Sentieon 202308.

Then, the relative mRNA amount per genome of yeast in a given growth condition is then calculated as:

$$\text{mRNA amount per genome} = \left( \frac{\sum \text{TPM}_{S.\ cerevisiae}}{\sum \text{TPM}_{C.glabrata}} \right) \Big/ \left( \frac{\sum \text{gDNA reads}_{S.\ cerevisiae}}{\sum \text{gDNA reads}_{C.glabrata}} \right),$$

which is used for calculating total mRNA concentration (see **Methods**).

### Cell size measurements

Cells were cultured in different growth conditions at 30 °C to an OD600 of 0.1–0.2. Cultures were briefly sonicated to disperse cell clumps and then diluted 1:100 in Isoton II diluent (Beckman Coulter, #8546719). Cell volumes were measured on a Beckman

Coulter Z2 counter following the manufacturer's protocol. Measurements <10 fL or >300 fL were excluded from downstream analysis.

### DNA content analysis

Four hundred μL of yeast culture in different growth conditions were fixed by adding 1 mL ice-cold 100% ethanol and incubating at 4 °C for 30 min. Fixed cells were pelleted (3,000 × g, 5 min, 4 °C) and washed twice with PBS. Pellets were resuspended in 50 mM sodium citrate (pH 7.2), treated with RNase A (0.2 mg/mL) at 37 °C overnight, then with proteinase K (0.4 mg/mL) at 50 °C for 1–2 h. Samples were stained with 16 μg/mL propidium iodide (Invitrogen) for 30 min at room temperature in the dark, sonicated briefly to disperse clumps, and analyzed on an Attune NxT flow cytometer (YL2-A channel), collecting >30,000 events per sample. Single-cell events were gated in FlowJo to exclude debris and aggregates, and the mean YL2-A fluorescence intensity was taken as the genomes per cell.

### Calculation of total mRNA concentration

To calculate the total mRNA concentration, we measured the relative mRNA amount per genome by spike-in RNA sequencing, the mean cell volume by Coulter counter, and genomes per cell by flow cytometry (see **Methods** described above). Relative mRNA amount per cell was then calculated as:

$$\text{mRNA amount per cell} = (\text{mRNA amount per genome}) \times (\text{genomes per cell}).$$

Then, total mRNA concentration was obtained by dividing mRNA amount per cell by the mean cell volume:

$$\text{total mRNA concentration} = \frac{\text{mRNA amount per cell}}{\text{mean cell volume}}.$$

**Extended Data Fig. 5b** shows a schematic of this workflow for quantifying total mRNA concentration.

### Flow cytometry measurements of global protein concentration

Yeast cultures were grown to an OD600 of 0.1, and 500 μL of culture was mixed with 500 μL of ice-cold 70% ethanol and incubated at 4 °C for 30 min to fix the cells. Fixed cells were washed twice with PBS and then resuspended in PBS containing 1 μg/mL Alexa Fluor 647 NHS Ester. Cells were stained for 1 h at room temperature in the dark, followed by three washes in PBS and brief sonication to disperse aggregates. Stained cells (>50,000 events per sample) were analyzed on an Attune NxT flow cytometer. Total protein fluorescence was recorded in the BL1-A channel, and forward scatter area (FSC-A) was used as an estimate of cell size. A robust linear regression of BL1-A versus FSC-A was performed for each sample, and the slope of the fitted line was taken as the measure of total protein concentration per unit cell volume.

### SLAM-seq to determine the global mRNA decay rate

This protocol was adapted from Alalam et al. with minor modifications [54]. Yeast cells were grown overnight in SC medium containing low uracil (0.1 mM) to the exponential phase ($OD_{600}$ ≈ 0.2) under the indicated carbon source conditions. Subsequently, 0.2 mM 4-thiouracil (4tU, sigma) was added to the culture, and cells were incubated for an additional 2 hours to allow metabolic labeling. A parallel culture treated with 0.2 mM DMSO was included as a negative control. For pulse–chase experiments, 50 mL of 4tU-labeled culture was rapidly harvested by vacuum filtration (GenClone) and immediately resuspended in 50 mL SC medium containing high uracil (20 mM) to initiate the chase. At defined time points (2, 7, 12, 17, and 22 minutes) after resuspension, 5 mL aliquots were collected and quenched by addition to pre-chilled 95% ethanol maintained on dry ice. In addition, one negative control sample was collected. Cells were pelleted by centrifugation and stored at −80 °C. Two biological replicates were performed for each carbon source condition.

To extract total RNA, cell pellets were lysed in 500 µL TRI Reagent using a FastPrep-24 instrument (5.5 m $s^{-1}$ for 30 s at 4 °C). RNA samples were protected from light prior to alkylation. Following lysis, 100 µL chloroform:isoamyl alcohol (24:1) was added to the supernatant, and samples were centrifuged at 16,000 × g for 15 min at 4 °C. The aqueous phase (~200 µL) was transferred to a new tube and mixed with 200 µL isopropanol containing 0.2 mM DTT and 1 µL glycogen as a carrier. Samples were incubated at room temperature for 10 min and centrifuged at 16,000 × g for 20 min at 4 °C to pellet RNA. The supernatant was discarded, and the pellet was washed with 500 µL of 75% ethanol containing 0.1 mM DTT. After centrifugation at 7,500 × g for 5 min, the pellet was air-dried for 10 min and resuspended in 20 µL nuclease-free water supplemented with 1 mM DTT. For alkylation, 1 µg total RNA was diluted in 15 µL water and mixed with 5 µL iodoacetamide (IAA; 100 mM in ethanol), 5 µL sodium phosphate buffer (500 mM, pH 8.0), and 25 µL DMSO. The reaction was incubated at 50 °C for 15 min and quenched by addition of 1 µL 1 M DTT. RNA was then precipitated by ethanol precipitation and resuspended in 10 µL nuclease-free water.

Quality control, library preparation, sequencing, and initial data processing of alkylated total RNA were performed by Lexogen NGS Services (Lexogen GmbH, Austria) using the SLAMdunk pipeline. Gene-level thymine-to-cytosine (T>C) conversion rates were calculated for downstream analysis. Due to the low total mRNA concentration in trehalose, genes with fewer than 10 counts per million (CPM) were excluded. In addition, samples with initial T>C conversion rates below 0.5% were also discarded. For all samples, background conversion rates were corrected by subtracting the corresponding negative control. The natural logarithm of T>C conversion rates was taken, and linear regression was performed over time, with the slope representing the mRNA decay rate. Genes with poor fitting quality ($R^2 < 0.5$) were excluded from further analysis.

## RT-qPCR to determine the mRNA decay rate

Yeast cells were grown in methionine-free SC medium with one of four carbon sources (glucose, galactose, glycerol, or trehalose) at 30 °C to mid-log phase. To shut off MET3/MET17 transcription, methionine was added to a final concentration of 1 mM [70]. At 0, 6, 10, 14, 18, and 22 min after addition, 1 mL culture aliquots were harvested by centrifugation (17,000 × g, 30 s), immediately snap-frozen in liquid nitrogen, and stored at –80 °C.

Cell pellets were lysed in 300 µL TRI Reagent (Zymo Research) using the FastPrep-24 (5.5 m/s for 30 s at 4 °C). Lysates were clarified (13,000 rpm, 2 min, 4 °C) and total RNA purified with the Direct-zol RNA Microprep Kit (Zymo Research, #R2061). One microgram of RNA was reverse-transcribed with SuperScript VILO Master Mix (Thermo Fisher), and cDNA was diluted 1:100 for quantitative PCR. RT-qPCR was performed using iTaq Universal SYBR Green Supermix (Bio-Rad, #1725121) with *ACT1* as the internal control.

For each time course, two biological replicates were assayed, each with two technical replicates. Standard $\Delta C_t$ values were plotted against time, and the mRNA decay rate was obtained by multiplying the fitted slope by ln(2).

## Nuclear-to-cell volume fraction measurements

Yeast cells expressing Pus1-eGFP were grown in SC medium with different nutrients at 30 °C to log phase (OD600=0.1-0.2). Cells were harvested by brief centrifugation (2,500 rpm, 3 min), and the pellet was gently resuspended and mounted on an agarose slide (see single molecule tracking section for a detailed description of the agarose slide).

Imaging was performed on a Zeiss Axio Observer.Z1 wide-field epifluorescence microscope equipped with a 63×/1.4 NA oil-immersion objective and a Colibri LED module. eGFP fluorescence (Pus1-eGFP) was excited using the 505 nm LED at 25% intensity with a 1 s exposure. Phase-contrast images were captured to delineate cell outlines. A TempController 37-2 unit maintained the sample at 30 °C throughout acquisition. For each condition, z-stacks were collected, and the plane with maximum focus was selected by eye. Over 500 cells were imaged per replicate.

Cell boundaries were segmented automatically using ACDC software [71] and manually calibrated, and nuclear regions were defined via a custom Python algorithm from fluorescence images. To calculate cell volume $V_{cell}$, each yeast cell was approximated as a prolate spheroid. Cross-sections perpendicular to the long axis were summed, and for budding cells the mother and bud volumes were computed separately and then combined. The nucleus was modeled as a tri-axial ellipsoid: the long (a) and short (b) axes were measured in 2D using the skimage package, the third axis (c) was set to $c=\frac{a+b}{2}$, and volume was calculated as $V_{nucleus}=\frac{4}{3}\pi abc$. For each cell, the nuclear-to-cell volume fraction was calculated as nuclear volume divided by cell volume.

## mRNA transcription model

Our model for mRNA dynamics is based on mass action kinetics of free nuclear RNAP II and the genome coupled with first order decay of mRNA at a rate that was found to be independent of the growth conditions (**Fig. 4e-g**). The DNA-bound RNAP II concentration, $[pol_{bound}]$, in the nucleus is determined by mass action with the genome, $[DNA]$, and the free nuclear RNAP II, $[pol_{free}]$, so that:

$$\frac{\mathrm{d}[pol_{bound}]}{\mathrm{d}t}=k_{on}^{pol}[DNA][pol_{free}]-k_{off}^{pol}[pol_{bound}],$$

where $k_{on}^{pol}$ and $k_{off}^{pol}$ are the binding and unbinding rates, respectively. At steady-state, we see that

$$k_{on}^{pol}[DNA][pol_{free}]-k_{off}^{pol}[pol_{bound}]=0.$$

We consider the fraction of RNAP II in the proteome, $\phi_{pol}$, to be equal to the sum of all subunits, so the concentration of RNAPII in nucleus is given as

$$[pol_{free}]+[pol_{bound}]=[pol_{total}]=P\phi_{pol}\frac{V_{cell}}{V_{nucleus}},$$

where $P$ is the global protein concentration, $V_{cell}$ is the cell volume, and $V_{nucleus}$ is the nuclear volume. We denote the ratio of nuclear-to-cell volume as $f={V_{nucleus}}/{V_{cell}}$. Substituting this into the above equation, the nuclear DNA-bound RNAP II concentration at steady-state can be written as:

$$[pol_{bound}] = \frac{P\phi_{pol}}{f\left(1+\frac{k_{off}^{pol}}{k_{on}^{pol}[DNA]}\right)}.$$

Since the DNA concentration in the nucleus equals the DNA amount divided by the nuclear volume, $[DNA] = \frac{DNA}{V_{nucleus}}$, we can rewrite the above equation as:

$$[pol_{bound}] = \frac{P\phi_{pol}}{f\left(1+\frac{K_d^{pol} V_{cell} f}{DNA}\right)},$$

where $K_d^{pol} = k_{off}^{pol} / k_{on}^{pol}$ is the dissociation constant of RNAPII and the genome. The fraction of DNA-bound RNAPII is:

$$\frac{[pol_{bound}]}{[pol_{total}]} = \frac{1}{1+\frac{K_d^{pol} V_{cell} f}{DNA}}.$$

For each growth condition, we can determine the cell volume with a Coulter counter (**Extended Data Fig. 5d**), and the nuclear-to-cell volume ratio with the Pus1-eGFP nuclear marker and a fluorescence microscope (**Fig. 4i**). Assuming the average DNA content in G1 phase is normalized to 1, the DNA content in different growth conditions can be determined by flow cytometry (**Extended Data Fig. 5c**). Thus, we can fit the model results with the single-molecule tracking data using only one free parameter $K_d^{pol}$. Fitting was performed using nonlinear least-squares regression in MATLAB (fitnlm), with 90% confidence intervals estimated from uncertainty in the fitted parameters propagated to the model predictions.

From the level of DNA-bound RNAPII and the constant mRNA decay rate, we can estimate the cellular mRNA concentration. mRNA is produced by RNAP II transcription in the nucleus at a rate $k_{trans}^{mRNA}[pol_{bound}]$, quickly exported to cytoplasm, and then degraded at a rate $k_d^{mRNA}$. Assuming that the timescale of mRNA production and degradation is much faster than that of cell growth, the system can be treated as being in steady state so that:

$$\frac{\mathrm{d}[mRNA]}{\mathrm{d}t} = k_{trans}^{mRNA}[pol_{bound}]f - k_d^{mRNA}[mRNA] = 0,$$

which can be solved for the mRNA concentration to yield:

$$[mRNA]=\frac{k_{trans}^{mRNA}P\phi_{pol}}{k_d^{mRNA}\left(1+\frac{K_d^{pol}V_{cell}f}{DNA}\right)}.$$

Here, global protein concentration (**Extended Data Fig. 5f**) and RNAP II fraction in proteome (**Fig. 4c**) were all measured. The global mRNA decay rate can be considered invariant across different carbon sources (**Fig. 4d-g**). Based on the fitted dissociation constant $K_d^{pol}$, we estimated the mRNA production rate $k_{trans}^{mRNA}$ by fitting the model using our measurements of the total mRNA concentration (**Fig. 4a**). Fitting was performed using nonlinear least-squares regression in MATLAB (fitnlm), with 90% confidence intervals estimated from uncertainty in the fitted parameters propagated to the model predictions.

## mRNA-ribosome translation model

Given the ribosome and mRNA concentrations as inputs, we construct a mass action kinetic model for protein translation that predicts the active ribosome fraction and cell growth rate. For simplicity, we ignore the assembly process of the 40S and 60S subunits and simply view the ribosome as being a single molecular complex. The translation process is simplified to 3 steps: ribosome binding, translation, and termination. The concentration of ribosomes bound at the initiation site is $[ribo_{init}]$ and the concentration of translating ribosomes is $[ribo_{trans}]$ so that:

$$\frac{d[ribo_{init}]}{dt}=k_{on}^{ribo}[RBS_{free}][ribo_{free}]-k_{init}^{ribo}[ribo_{init}] \text{ and}$$

$$\frac{d[ribo_{trans}]}{dt}=k_{init}^{ribo}[ribo_{init}]-k_{off}^{ribo}[ribo_{trans}],$$

where free ribosomes, $[ribo_{free}]$, bind the initiation site, $[RBS_{free}]$, via mass action kinetics at a rate $k_{on}^{ribo}$. The rate of initiating translation is $k_{init}^{ribo}$ and the termination rate is $k_{off}^{ribo}$.

In addition, we can write the total ribosome concentration, $[ribo_{total}]$, as the sum of the distinct pools of ribosomes so that

$$[ribo_{total}]=[ribo_{free}]+[ribo_{init}]+[ribo_{trans}].$$

Finally, we use the fact that the mRNA concentration equals to concentration of free ribosome binding sites (only one per mRNA) and the concentration of occupied ribosome binding sites:

$$[mRNA_{total}] = [ribo_{init}] + [RBS_{free}].$$

We then make the above equations dimensionless by scaling our variables. *x* denotes the translating ribosome fraction and *y* denotes the initiating ribosome fraction of total ribosome. The dimensionless equations are:

$$\frac{dx}{d\tau} = \gamma y - x$$

$$\frac{dy}{d\tau} = (\alpha - \beta y)(1 - x - y) - \gamma y.$$

where the dimensionless parameters are given by:

| Parameters | |
|---|---|
| $\alpha = \frac{[mRNA] k_{on}^{ribo}}{k_{off}^{ribo}}$ | Relative mRNA concentration |
| $\beta = \frac{[ribo_{total}] k_{on}^{ribo}}{k_{off}^{ribo}}$ | Relative ribosome concentration |
| $\gamma = \frac{k_{init}^{ribo}}{k_{off}^{ribo}}$ | Relative initiation rate |
| $\tau = t \times k_{off}^{ribo}$ | Dimensionless time |

At steady-state:

$$\gamma y - x = 0,$$

$$(\alpha - \beta y)(1 - x - y) - \gamma y = 0.$$

Thus, we have $x_{ss} = \gamma y_{ss}$. Replacing the above equation gives

$$(\alpha - \beta y)(1 - \gamma y - y) - \gamma y = 0,$$

Which can be rewritten as

$$(\gamma\beta + \beta) y^2 - (\beta + \alpha\gamma + \alpha + \gamma) y + \alpha = 0.$$

We note that $\alpha$, $\beta$ and $\gamma > 0$, and the translating ribosome fraction should also be positive. The above equation can then be solved to calculate the translating ribosome fraction $y$ as a function of the relative mRNA and ribosome concentrations, $\alpha$ and $\beta$, respectively, so that:

$$y = \frac{\alpha + \beta + \gamma + \alpha\gamma - \sqrt{(\alpha + \beta + \gamma + \alpha\gamma)^2 - 4\alpha\beta(\gamma + 1)}}{2\beta(\gamma + 1)}.$$

We define the active ribosome fraction $\phi_a = x + y$ so that the above equation can be rearranged to yield:

$$\phi_a = \frac{\alpha + \beta + \gamma + \alpha\gamma - \sqrt{(\alpha + \beta + \gamma + \alpha\gamma)^2 - 4\alpha\beta(\gamma + 1)}}{2\beta}.$$

We next evaluate three dimensionless parameters as follows. The parameter $\alpha_{glc}$ is defined as a dimensionless mRNA concentration in glucose, which allows experimentally measured relative changes in ribosomal protein and mRNA levels to be converted into dimensionless concentrations across growth conditions. Yeast cells typically contain ~60,000 mRNA transcripts [31,72] and ~ 180,000-300,000 ribosomes [32,72] in glucose, so that the ribosome-to-mRNA ratio $\beta_{glc}/\alpha_{glc}$ ~3. We use a linear fit to our data to determine the relationship between both ribosome and mRNA concentrations and the growth rate so that the ribosome-to-mRNA ratios $\beta/\alpha$ at different growth rates can be calculated. $\gamma$, the initiation rate-to-termination rate ratio can be estimated from the number of initiating and elongating ribosomes. Approximately 4% of bound ribosomes (2% in polysomes, 7% in monosomes) are engaged in initiation in glucose [26], thus we set $\frac{x_{glc}}{y_{glc}} = \gamma_{glc} = 24$. Our results show that the elongation rate remains constant across the tested carbon sources. Here, we assume the initiation rate is also constant and infer an initiation-to-termination ratio ($\gamma$) of about 24 in all conditions. Then, the active ribosome fraction is a function of the growth rate with only one free parameter $\alpha_{glc}$, whose best fit value is 2.05 (least-squares fit using MATLAB's fitnlm), and 90% confidence intervals were obtained by propagating uncertainty in the fitted parameter to the model predictions. After determining the fraction of active ribosomes from mRNA and ribosome concentrations, the cellular growth rate can be estimated using the method described in the section "Estimate average peptide elongation rate of active ribosomes", setting the peptide elongation rate to 9 amino acids per second.

**G1 arrest experiment**

G1 arrest time-course experiments were done using SCGE media. These strains lack the endogenous G1 cyclins (*cln1Δcln2Δcln3Δ*) require β-estradiol (BE) to express *CLN1* from a synthetic promoter. When this strain was proliferating, we used a concentration of 20 nM BE (Sigma-Aldrich, cat. no. E2758). To trigger the arrest, cells were washed and resuspended in pre-warmed medium lacking BE at t = 0 minutes. We measure the growth rate by both Beckman Z2 Coulter counter and spectrophotometer every 20 min. At t minutes, the local growth rate was determined by calculating the slope of the cell size/OD versus time using measurements taken at five points: t–40 min, t–20 min, t, t+20 min, and t+40 min. We then take the average of local growth rates measured by OD600 and cell size.

To estimate the active ribosome fraction and growth rate in **Fig. 6c**, we introduced mRNA concentration data in SCGE after G1 cyclin shut-off from Swaffer et al. [25], where the mRNA concentration in SCGE before arrest is estimated as 0.4 relative to glucose. Since the total ribosome content after G1 arrest is constant [58], we set ribosome content as 15.5% of the proteome as determined in **Fig. 1a**.

### Fluorescence in situ hybridization (FISH) to determine total mRNA concentration

Yeast were grown to an OD600 of 0.2–0.4. Forty-five mL of culture were fixed by adding 5 mL 37% formaldehyde and incubating 45 min at room temperature. Cells were centrifuged at 1,600 × g for 4 min, resuspended in 1 mL ice-cold Fixation Buffer (1.2 M sorbitol, 0.1 M $K_2HPO_4$, pH 7.5), and washed once with the same buffer. Cells were then resuspended in 1 mL Fixation Buffer containing zymolyase and incubated at 30 °C for 90 min. After two washes (400 × g, 5 min each), cells were permeabilized in 70% ethanol at 2–8 °C overnight.

For hybridization, ethanol-stored cells were collected (400 × g, 5 min) and resuspended in 100 µL hybridization mix containing 125 nM probes (Stellaris T30-Quasar 570) in Stellaris Hybridization Buffer with 10% formamide. Samples were incubated overnight at 30 °C in the dark. Post-hybridization, 100 µL Stellaris Wash Buffer A was added and the samples were centrifuges at 400 × g for 5 min, resuspended in 1 mL Wash Buffer A and incubatee 30 min at 30 °C in the dark. The samples were then centrifuged again (400 × g, 5 min), resuspended in 1 mL Stellaris Wash Buffer B and incubated 2–5 min at room temperature before being centrifuged once more (400 × g, 5 min). Cells were then resuspended in 15 µL Vectashield mounting medium. 5 µL of the sample was mounted on a glass slide, covered with an 18 × 18 mm coverslip, and sealed at the edges with nail polish. Imaging was performed on a Zeiss Axio Observer.Z1 wide-field epifluorescence microscope with a 63×/1.4 NA oil-immersion objective and a Colibri LED module. FISH signals were acquired using the 555 nm LED at 100% intensity with a 1 s exposure, and phase-contrast images were captured to delineate cell outlines.

Cell boundaries were segmented automatically using ACDC software and then checked manually [71]. After background subtraction, T30 FISH fluorescence concentration was calculated as the sum of cellular fluorescence divided by cell volume.

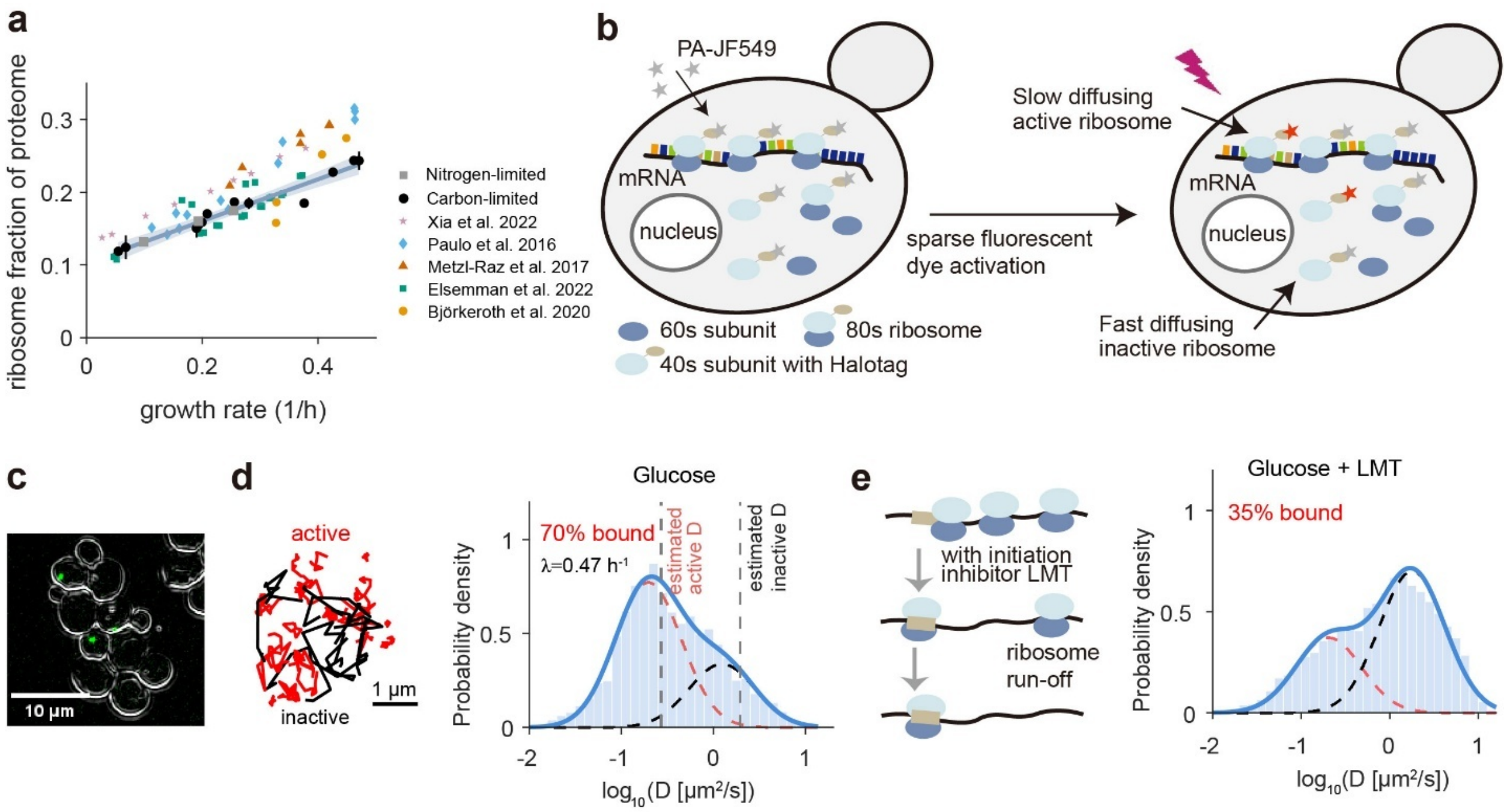

**Figure 1: Quantification of active ribosome fraction using single-molecule tracking (SMT).**

**a**, Total ribosome fraction of the proteome increases with growth rate. Black dots are our data, while colored dots are data from previous studies [3–7]. The blue line is the best-fit linear regression with 99% confidence interval (CI). Data are presented as mean ± SEM (n = 3 for glucose; n = 2 for galactose, glycerol, and trehalose).

**b**, Schematic diagram illustrating single-molecule tracking of ribosomes. Active ribosomes bound to mRNA exhibit slower diffusion compared to free, inactive ribosomes.

**c**, Composite bright-field and fluorescence single-molecule tracking image. Green dots indicate individual ribosomes.

**d**, Left: Representative trajectories of active (red) and inactive (black) ribosomes as measured by single-molecule tracking. Right: The fractions of active and inactive ribosomes were distinguished using a Gaussian mixture model. Distribution of ribosomal diffusion coefficients in yeast cultured in glucose (n = 4,091 tracks). Gray dotted lines indicate theoretical estimates of the diffusion coefficients of active and inactive ribosomes.

**e**, Left: Schematic depicting lactimidomycin (LMT) treatment, which inhibits ribosomes at the translation initiation site. Right: Distribution of ribosomal diffusion coefficients following 4 µM LMT treatment (n = 5,332 tracks).

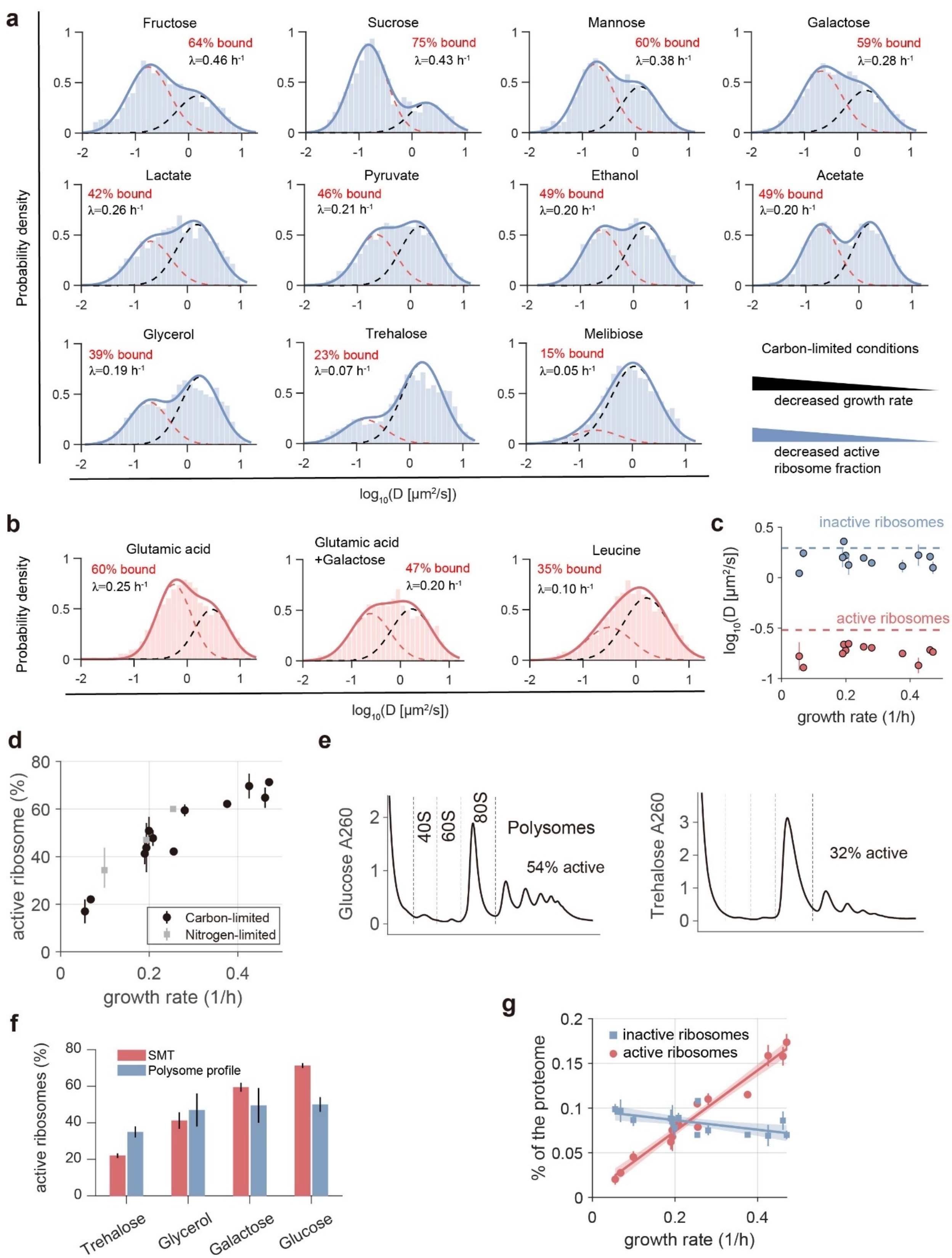

**Figure 2: The fraction of active ribosome is positively correlated with growth rate.**

**a**, Distributions of ribosomal diffusion coefficients across yeast cultured in SC media with fructose (n=1,916 tracks), sucrose (n=2,376 tracks), mannose (n=3,730 tracks), acetate (n=2,790 tracks), galactose (n=4,679 tracks), lactate (n=2,269 tracks), ethanol (n=4,231 tracks), pyruvate (n=1,881 tracks), glycerol (n=2,548 tracks), trehalose (n=3,351 tracks), and melibiose (n=3,722 tracks).

**b**, Distributions of ribosomal diffusion coefficients for yeast cultured in nitrogen-limited conditions with glutamic acid (n=5,091 tracks), leucine (n=3,853 tracks) and glutamic acid+galactose combination (n=3,422 tracks).

**c**, Average diffusion coefficient of active (red) and inactive (blue) ribosomes in 12 carbon-limited conditions. Dotted lines indicate the predicted diffusion coefficients, as described in **Fig. 1d**.

**d**, The fraction of active ribosomes increases with growth rate. Black dots represent carbon-limited conditions, while gray dots represent nitrogen-limited conditions. The fraction of active ribosomes is shown as mean ± range (n=2), while growth rate is shown as the mean of three biological replicates (n=3).

**e**, Polysome profiling results for yeast cells cultured in SC medium with glucose and trehalose. Peaks corresponding to 40S, 60S, monosomes, and polysomes are indicated.

**f**, Comparison of the fraction of active ribosomes measured by single-molecule tracking (red) and polysome profiling (blue) across four conditions. Data are mean ± range (n = 2).

**g**, Variation of active and inactive ribosome fractions of the proteome with growth rate. Active ribosome content was calculated as the product of total ribosome content and the active ribosome fraction. Data are presented as mean ± SEM (from error propagation). Red and blue lines represent linear fits of active and inactive ribosome content, respectively. Shaded areas indicate the 90% CI.

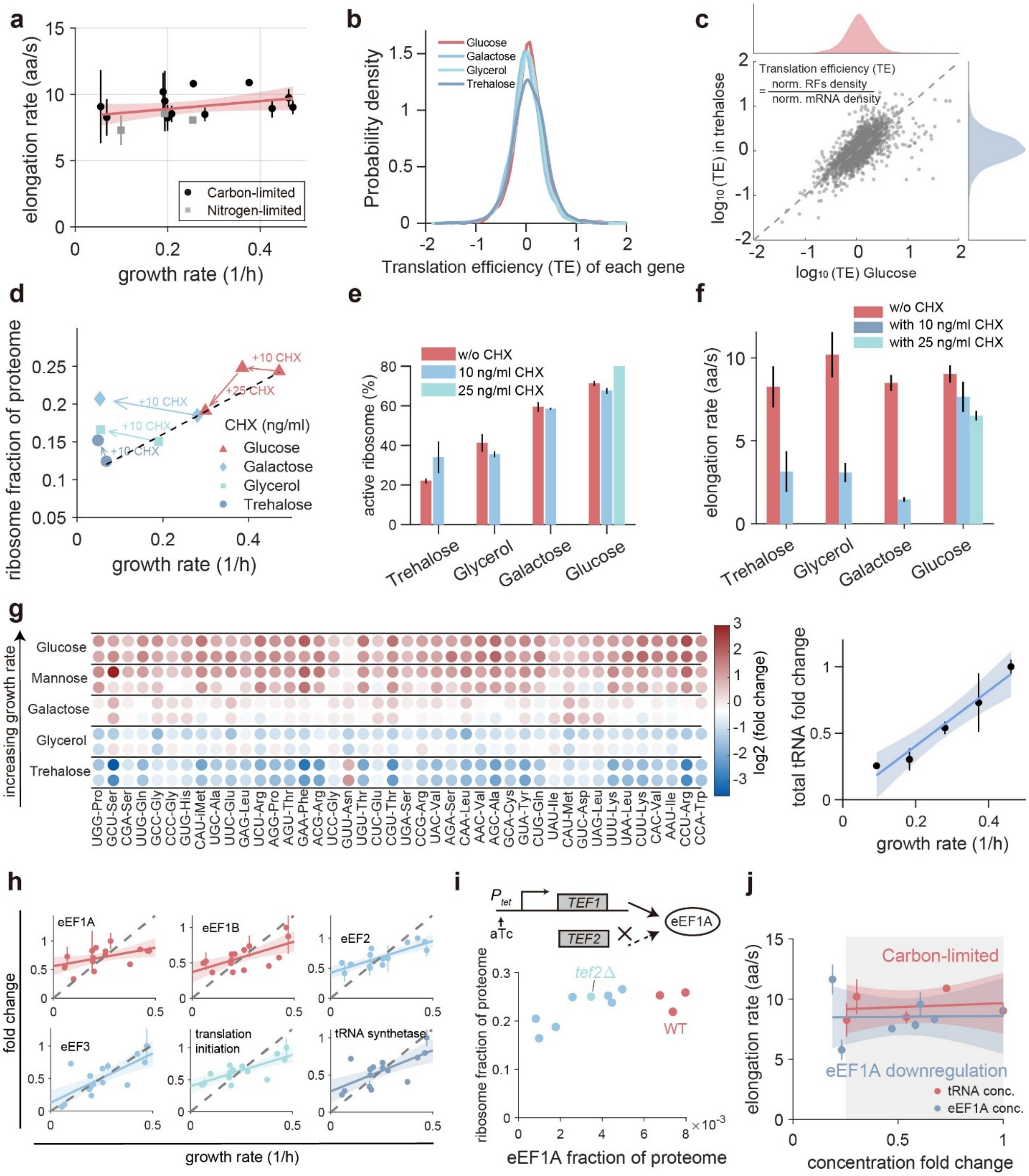

**Figure 3: Buffered tRNA maintains a constant peptide elongation rate across different growth conditions**

**a**, Peptide elongation rates calculated for cells growing on 12 different carbon sources. The red line represents a linear fit with a 90% confidence interval. Data are shown as mean ± SEM (from error propagation).

**b**, Distribution of translation efficiency (TE) of 2,055 genes under the 4 indicated growth n our model, the concentration of DNA-bound RNAP II, $[pol_{bound}]$, is determined by mass action kinetics of free RNAP II, $[pol_{free}]$, and the nuclear genome concentration, $[DNA]$. conditions. TE is defined as normalized ribosome footprint density divided by normalized mRNA density (see **Methods**).

**c**, Comparison of translation efficiency (TE) for 2,055 genes between glucose (fast growth) and trehalose (slow growth). Marginal histograms show the TE distributions under each carbon source.

**d**, Variation in ribosome content following treatment with the translation inhibitor cycloheximide (CHX) across four carbon sources. Colors indicate different carbon sources and arrows represent changes in ribosome content upon CHX treatment. The black dashed line shows a linear fit of ribosome content without CHX.

**e**, Distributions of ribosomal diffusion coefficients determine the active ribosome fraction with and without CHX treatment in indicated conditions. Data are shown as mean with range (n=2).

**f**, Peptide elongation rates calculated for cells growing on four carbon sources following CHX treatment. Gray bars represent elongation rates without CHX treatment for comparison. Data are shown as mean ± SEM (from error propagation).

**g**, The heatmap shows the $\log_2$ fold changes of all 42 tRNA concentrations measured by RT-qPCR. Values of each tRNA were normalized to the mean tRNA levels across the five carbon sources. The right panel displays the total tRNA concentration fold changes across these conditions. Data show the mean with range (n=2).

**h**, Fraction of the proteome plotted against growth rate for elongation factor eEF1A, eEF1B, eEF2, eEF3, tRNA synthetase, and all initiation factors added together (**Supplementary Table 2; Extended Data Fig. 4c**). Solid lines indicate linear fits with 90% CI. The dashed black line denotes proportional scaling with growth rate. Data are presented as mean ± range (n = 3 for glucose; n = 2 for galactose, glycerol, and trehalose) and normalized to the maximal value.

**i**, Schematic diagram of exogenously controlled eEF1A protein expression. The *TEF2* gene was deleted, and the promoter of *TEF1* was replaced with an anhydrotetracycline (aTc)-inducible $P_{tet}$ promoter. The bottom panel shows ribosome content at different eEF1A expression levels.

**j**, Red dots show the peptide elongation rate remains nearly constant across different carbon sources despite changes to the total tRNA concentration. Blue dots show the peptide elongation rate as a function of eEF1A expression levels in the $P_{tet}$-*TEF1* system (mean ± SEM, estimated by error propagation). Solid lines are linear fit with 90% confidence intervals, and the gray shaded area indicates the range in which tRNA is buffered.

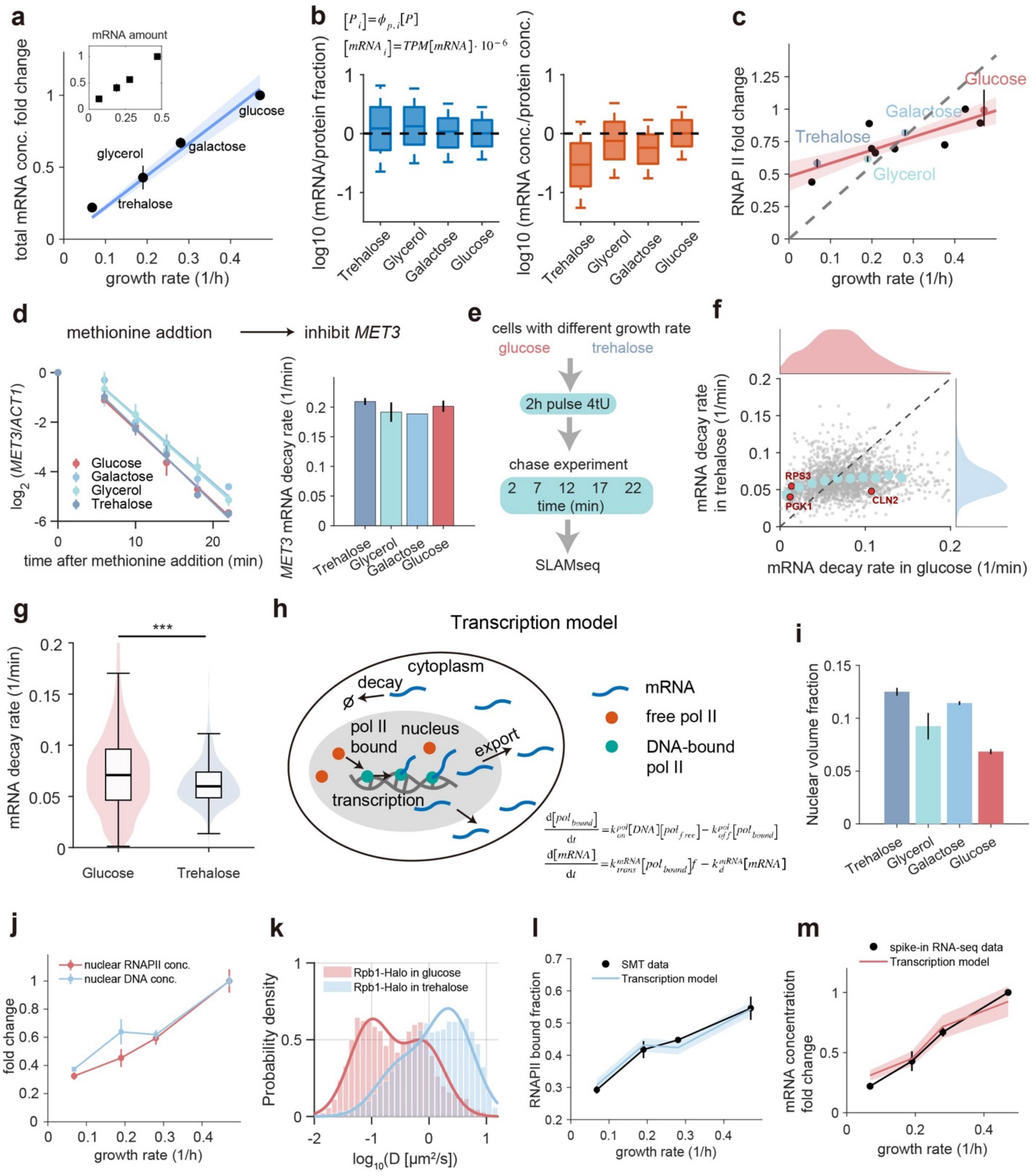

## Figure 4: RNAPII kinetics scale total mRNA concentration with growth rate

**a**, mRNA concentration measured by spike-in mRNA sequencing in four conditions (see **Extended Data Fig. 5b**). Data are mean with SEM (estimated from error propagation). The blue line denotes the linear fit of mRNA concentration with 90% CI. The inset shows the corresponding changes in mRNA amount.

**b**, Left: Median with interquartile range (IQR) of the mRNA fraction-to-protein fraction ratio for 1,795 genes across four growth conditions. Right: Median with IQR of mRNA concentration-to-protein concentration of 1,795 genes across four growth conditions. The individual protein concentrations were calculated by multiplying its protein fraction by the total protein concentration. The mRNA fraction for each gene is its TPM value divided by $10^6$. mRNA concentrations were calculated as the product of the mRNA fraction and total mRNA concentration.

**c**, RNAP II protein expression level across 12 different carbon sources, where four conditions were highlighted in which mRNA concentration was also measured. The shaded area denotes the 90% CI of a linear fit. Data are presented as mean ± range (n = 3 for glucose; n = 2 for galactose, glycerol, and trehalose).

**d**, Left: *MET3* mRNA concentrations relative to *ACT1* mRNA after methionine addition represses *MET3* transcription in cells growing using 4 different carbon sources. Right: Mean (± range) of *MET3* mRNA decay rates measured in two biological replicates.

**e**, Schematic diagram showing the experimental setup for SLAM-seq.

**f**, mRNA decay rates for each gene in glucose and trehalose conditions (N = 1,966), with three representative genes highlighted. The black dashed line represents y = x. Cyan dots denote binned averages. Marginal histograms show the distributions of decay rates in each carbon source.

**g**, Violin plots show the distribution across genes, with embedded box plots indicating the median and IQR (*N* = 1,966). Statistical significance was assessed using both the Kruskal-Wallis test (***, $p < 0.001$) and the paired Wilcoxon signed-rank test (***, $p < 0.001$).

**h**, Schematic diagram showing transcription driven by mass action kinetics of RNAP II and the genome.

**i**, Nuclear fraction of cell volumes measured by Pus1-eGFP fluorescent nuclear protein in four conditions. Data are presented as mean with range from two biological replicates (see **Extended Data Fig. 6f**).

**j**, Nuclear RNAP II and DNA concentrations in the indicated conditions. Data are mean with SEM (from error propagation).

**k**, Histogram of the diffusion coefficient of Rpb1, determined from single-molecule tracking, for *RPB1-Halo* in glucose (n = 3,944 tracks) and *RPB1-Halo* in trehalose (n=12,786 tracks).

**l**, The fraction of DNA-bound Rpb1 molecules as a function of growth rate. Data (black dots) are shown as the mean with range, n=2. The bound fraction of RNAPII is predicted by our transcription model. The shaded area indicates the 90% confidence interval of the model estimate (see **Methods**).

**m**, Predicted mRNA concentration changes from our transcription model. The shaded area indicates the 90% confidence intervals of the model estimate (see **Methods**).

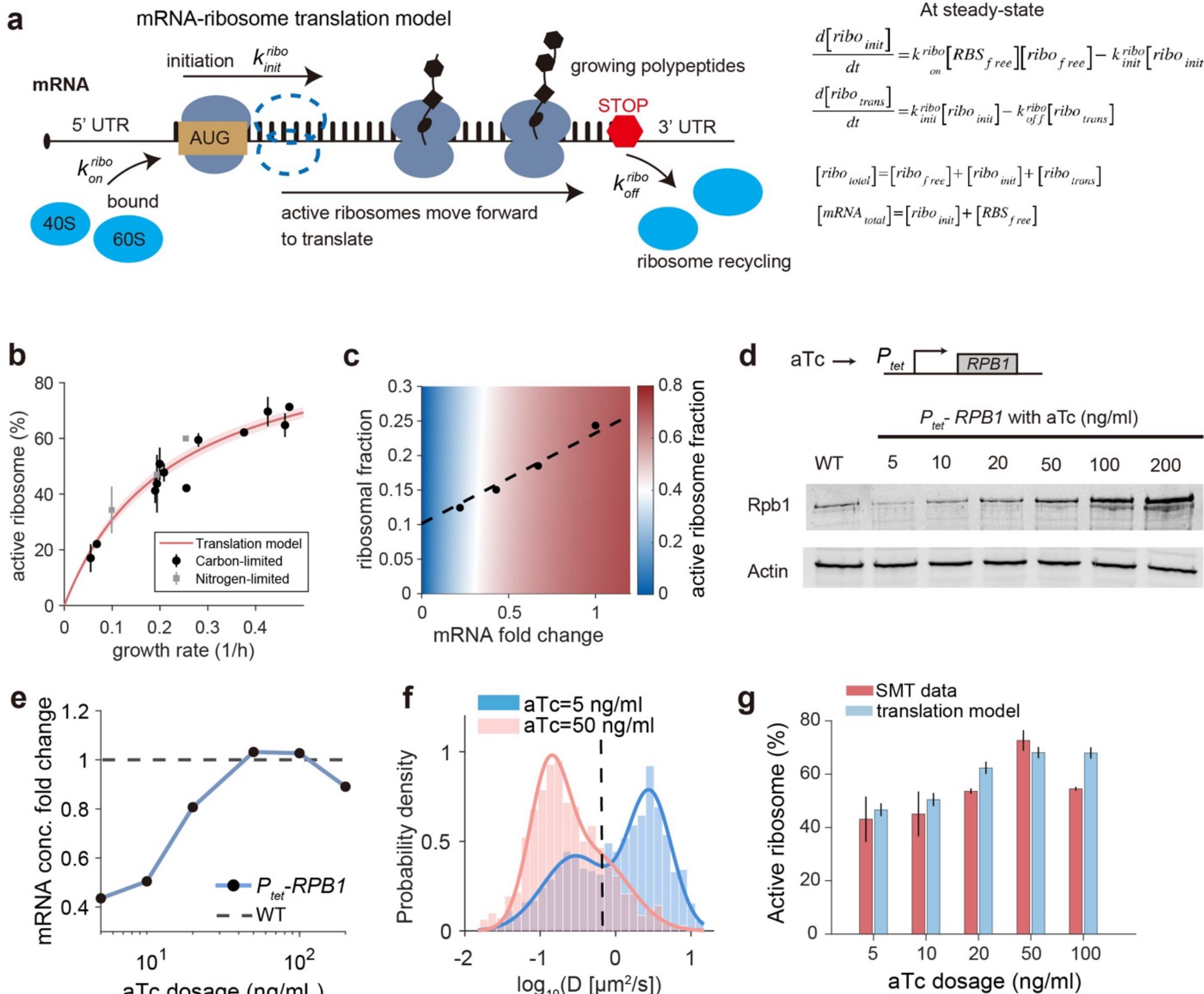

**Figure 5: mRNA and ribosome mass action kinetics determine the active ribosome fraction.**

**a**, Schematic illustrating the mRNA–ribosome mass action kinetic model for protein translation.

**b**, The fraction of active ribosomes were fit using our translation model. The shaded area represents the 90% confidence interval of the single parameter model fit (see **Methods**). Experimental data are from **Fig. 2c**.

**c**, Heatmap showing how relative mRNA and ribosome concentrations affect the fraction of active ribosomes (color scale). Data points represent measurements from four growth conditions. The yellow line indicates the coordinated adjustment of mRNA and ribosome levels with increasing growth rates.

**d**, Immunoblot showing Rpb1 protein expression at different aTc concentrations for a *$P_{tet}$-RPB1* strain.

**e**, The fold change of mRNA concentration of the $P_{tet}$-*RPB1* strain at the indicated aTc concentrations. mRNA concentrations were quantified by spike-in RNA sequencing. The dashed line indicates the total mRNA concentration of the wild-type strain grown in glucose.

**f**, Distribution of ribosome diffusion coefficients in a $P_{tet}$-*RPB1* strain grown in media with 5 (n=1,785) and 50 ng/ml (n=1,490) aTc concentrations.

**g**, The fraction of active ribosomes in the $P_{tet}$-*RPB1* strain with different aTc concentrations, measured by single-molecule tracking (SMT). Data are shown as mean ± SEM (n = 3 for 5 and 10 ng/mL aTc; n = 2 for all others) and compared with the predicted values of our translation model. The 90% confidence interval of predictions were obtained by propagating uncertainty in the fitted parameters estimated in **Fig. 5b**.

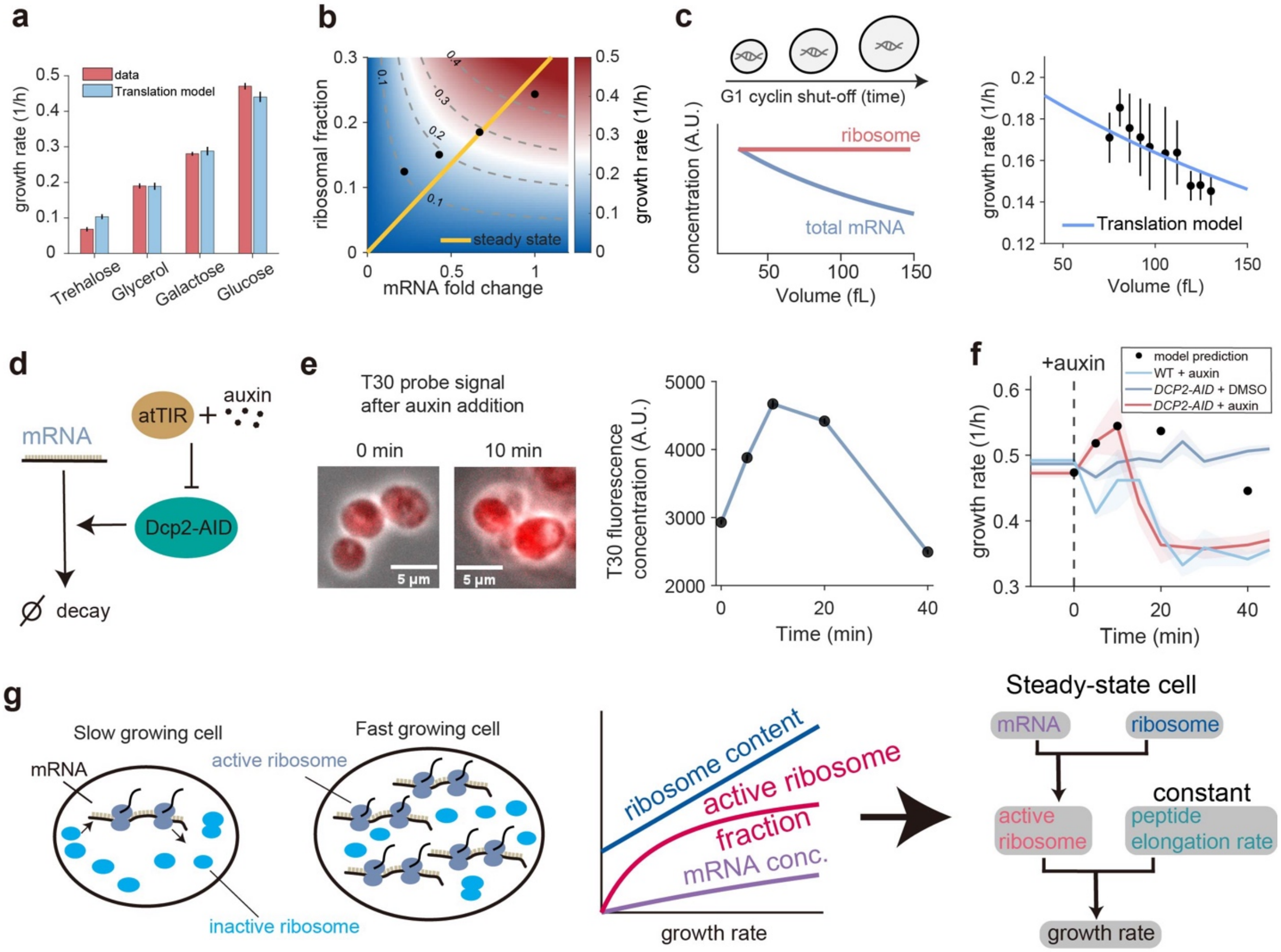

**Figure 6: Yeast cells coordinate mRNA and ribosome concentrations to determine growth rate.**

**a**, The growth rates in four conditions (mean with SEM, n=3) were compared with model results, which are calculated from the translation model taking the mRNA and ribosome concentrations as input. The 90% confidence interval of predictions were obtained by propagating uncertainty in the fitted parameters estimated in **Fig. 5b**.

**b**, Heatmap showing the effect of relative mRNA and ribosome concentrations on growth rate. Gray contour lines indicate equal growth rate. The yellow solid line indicates proportional increases of mRNA and ribosome concentrations.

**c**, Left panel: Schematic indicating G1 arrested yeast cells growing larger to dilute the genome and dilute mRNA while the ribosome concentration remains constant. Right: Comparison of the predicted and measured growth rates for these cells. Data show mean values with SEM (n=4).

**d**, Schematic diagram illustrating conditional degradation of Dcp2-AID upon auxin addition.

**e**, Left panel: Representative composite phase-contrast and fluorescence images showing T30 FISH signal at 0 and 10 minutes after auxin addition. Right panel: T30 FISH fluorescence concentration at the indicated times after auxin addition. Data represent mean ± SEM (n = 914 at 0 min; n=875 at 5 min; n=1,592 at 10 min; n=1,386 at 20 min; n=968 at 40 min).

**f**, Growth rate, estimated as the slope of $log_2$-transformed optical density (OD600) as a function of time using a centered 4-point sliding-window linear regression in the indicated conditions. The shaded area shows the SEM (n = 3), and the black dots mark the model prediction based on total mRNA concentration.

**g**, Schematic diagram illustrates how cell growth is organized in fast and slow growth conditions. Yeast cells in slow growth conditions have less mRNA and ribosome concentrations, while cell in fast growth conditions have more mRNA and ribosome conditions. The peptide elongation rate remains nearly constant across conditions. Mass-action kinetics between mRNA and ribosomes determine the concentration of active ribosomes, which further sets the growth rate.

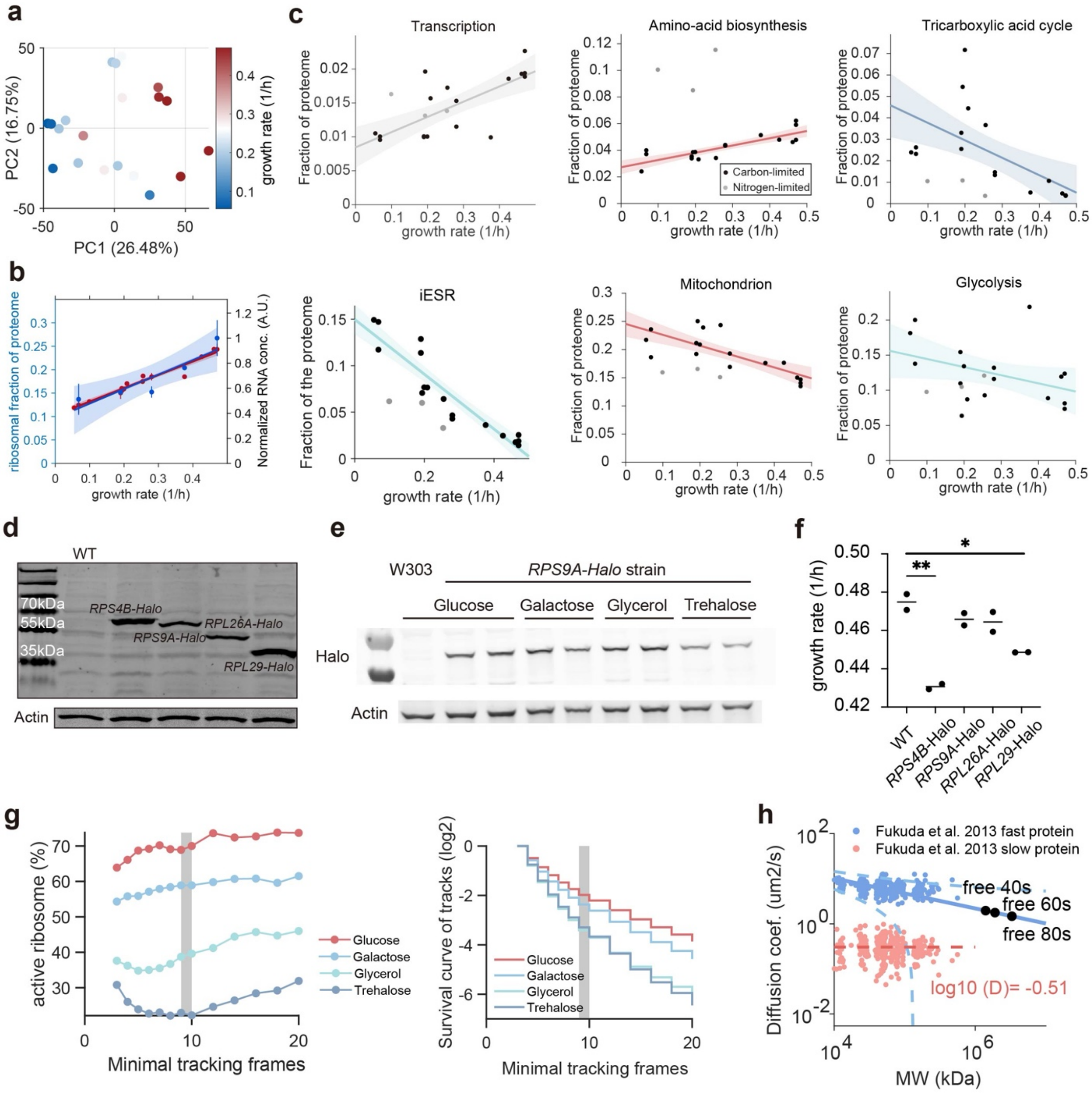

**Extended Data Fig. 1: Quantification of active ribosome fraction by single-molecule tracking (SMT), related to Fig. 1.**

**a**, Principal component analysis of proteomic data for yeast grown on the indicated carbon sources, where color denotes the growth rate.

**b,** Ribosomal protein fraction of the proteome quantified by mass spectrometry is shown on the left axis (red; data from **Fig. 1a**). Normalized total RNA concentration is shown on the right axis (blue, see **Methods**). Data are presented as mean with range (n = 2). Shaded regions indicate the 90% confidence interval.

**c,** Proteome fractions of proteins involved in amino acid biosynthesis, nucleotide biosynthesis, the TCA cycle, glycolysis, mitochondrial function, and lipid biosynthesis are shown as a function of growth rate across 15 growth conditions. Black dots indicate carbon-limited conditions, and grey dots indicate nitrogen-limited conditions. Solid lines represent linear fits, and shaded areas denote the 90% confidence intervals.

**d**, Immunoblot analysis confirms successful incorporation of a Halo tag into the ribosomal proteins Rps4B, Rps9A, Rpl26A, and Rpl29 in the 4 indicated strains.

**e**, Immunoblot showing the expression of the Halo tag in an *RPS9A-Halo* strain in the indicated SC media.

**f**, Growth rates for the indicated strains in SCD media. * denotes $p<0.05$ and ** denotes $p< 0.01$ calculated using an unpaired t-test was performed.

**g**, Left: The active ribosome fraction calculated by different minimum numbers of single molecule tracking frames. Right: Fraction of molecules tracked for more than the indicated number of frames. Measurements based on trajectories lasting at least 9 frames (shaded region) provided robust results balancing the tradeoff of getting more frames with the loss of molecules during tracking.

**h**, Diffusion coefficient of free ribosome (*D*) can be estimated using the Stokes-Einstein equation and is a function of molecular weight (MW) $D_i \sim k \times MW_i^{-\frac{1}{3}}$. The active ribosome diffusion coefficient was approximated from average values of other slow-moving proteins.

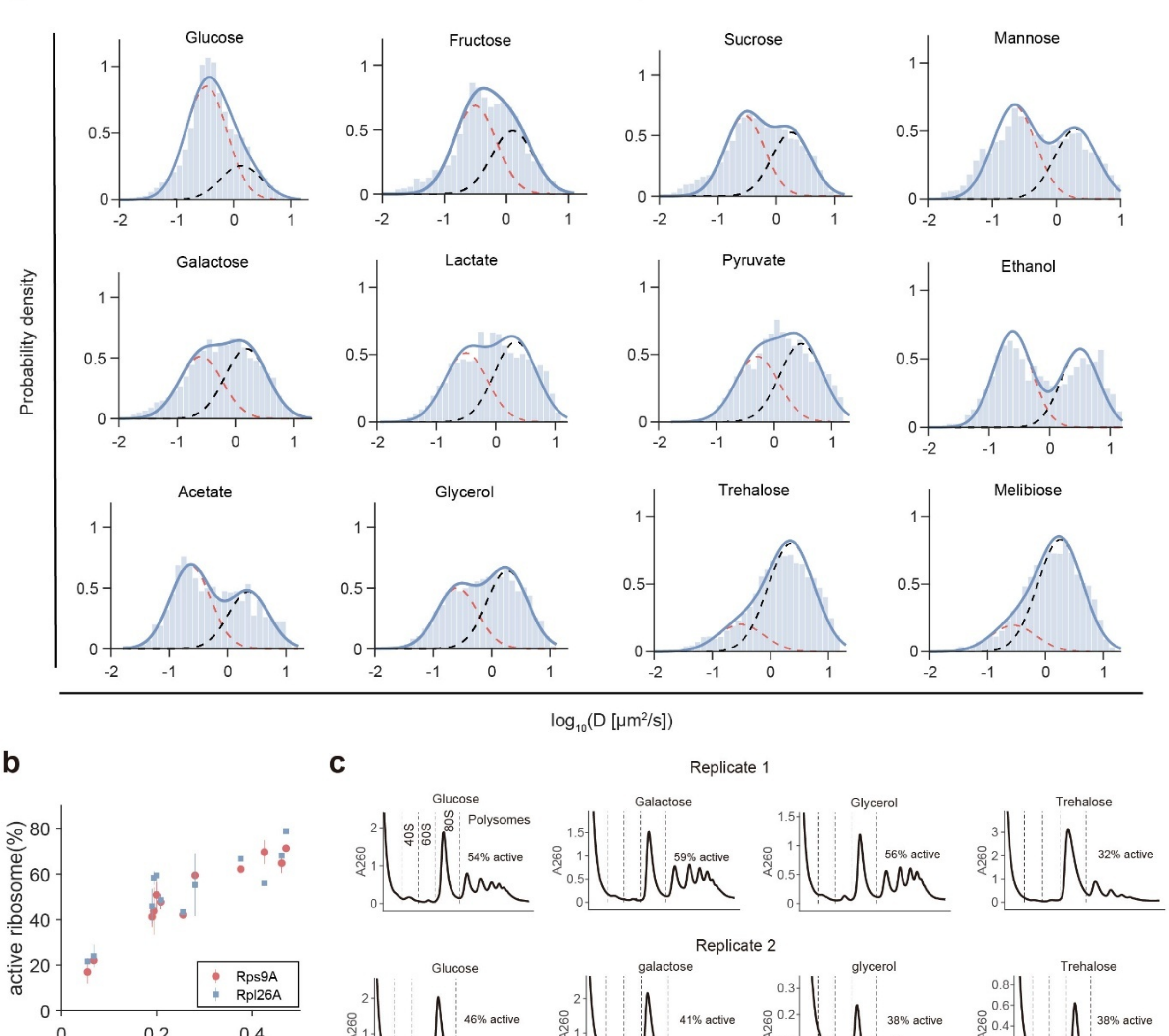

**Extended Data Fig. 2: Active ribosome fraction measured using *RPL26A-Halo* to tag the large subunit, related to Fig. 2.**

**a**, Diffusion coefficient distributions were determined using RPL26A-Halo for cells grown on indicated carbon sources. The number of tracks analyzed was n = 11,209 for glucose, n = 5,075 for fructose, n = 12,061 for sucrose, n = 1,806 for mannose, n = 1,078 for acetate, n = 6,008 for galactose, n = 2,306 for lactate, n = 1,518 for ethanol, n = 2,643 for pyruvate, n = 5,942 for trehalose, and n = 8,552 for melibiose.

**b**, Comparison of active ribosome fractions measured by tagging Rps9A (40S subunit) and Rpl26A (60S subunit) with Halo tags. Data are presented as mean with range (n=2).

**c**, Polysome profiling results for yeast cells cultured in SC medium with glucose, galactose, glycerol, and trehalose (n=2). Peaks corresponding to 40S, 60S, monosomes, and polysomes are indicated.

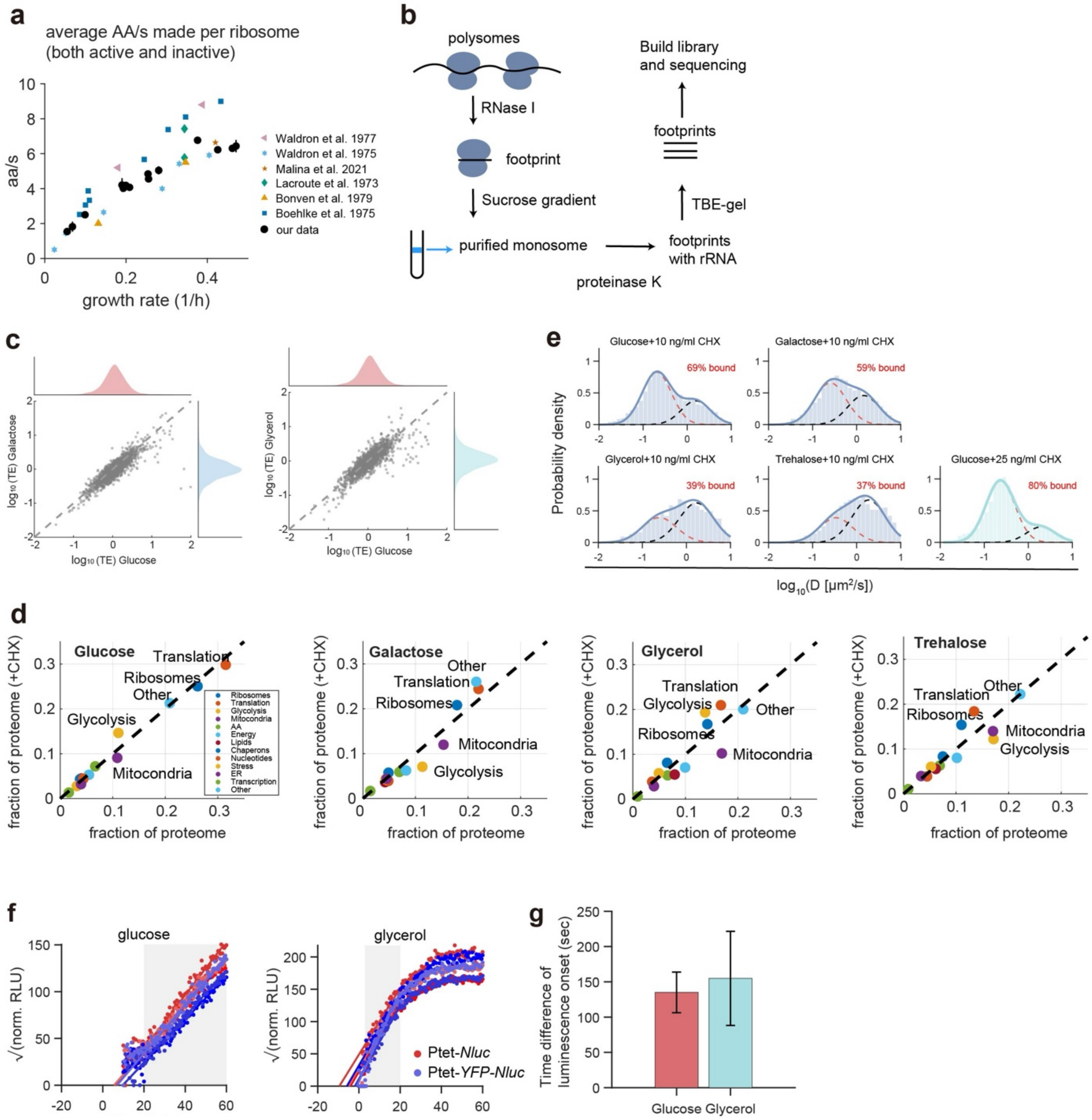

**Extended Data Fig. 3: Translation machinery and rate analysis, related to Fig. 3.**

**a**, Under the assumption that all ribosomes are active, the average elongation rate of total ribosomes is shown to monotonically increase with the growth rate. Colored dots indicate data from previous studies [17–21,41]. Data are presented as mean ± SEM (from error propagation).

**b**, Workflow schematic illustrating the procedure for ribosome footprint sequencing.

**c**, Comparison of translation efficiency (TE) for 2,055 genes under two pairs of conditions. The left panel contrasts glucose with galactose, and the right panel contrasts glucose with glycerol. Marginal histograms display the TE distributions for each condition.

**d**, Comparison of overall proteome composition after CHX treatment in four carbon sources. The proteomic classification follows categories defined by previous work[4].

**e**, Distributions of ribosomal diffusion coefficients (CHX-treated) used to estimate the active ribosome fraction in SC medium: glucose (+10 ng/mL CHX, n = 12,079; +25 ng/mL CHX, n = 4,243), galactose (+10 ng/mL CHX, n = 5,153), glycerol (+10 ng/mL CHX, n = 3,018), and trehalose (+10 ng/mL CHX, n = 4,407).

**f**, Luminescence was measured over time after aTc induction in two reporter constructs: $P_{tet}$-*Nluc* and $P_{tet}$-*YFP-Nluc*. The square root of normalized relative luminescence units (RLU) was plotted to estimate translation onset (see **Methods**). Linear fits within the shaded region were used to extrapolate the time of first detectable luminescence. Three biological replicates were performed for each condition.

**g**, Time differences of $P_{tet}$-*Nluc* and $P_{tet}$-*YFP-Nluc* luminescence onset between glucose and glycerol conditions. Error bars represent the SEM of the time difference between YFP-Nluc and Nluc onset, calculated from the 3×3 pairwise comparisons (n = 9).

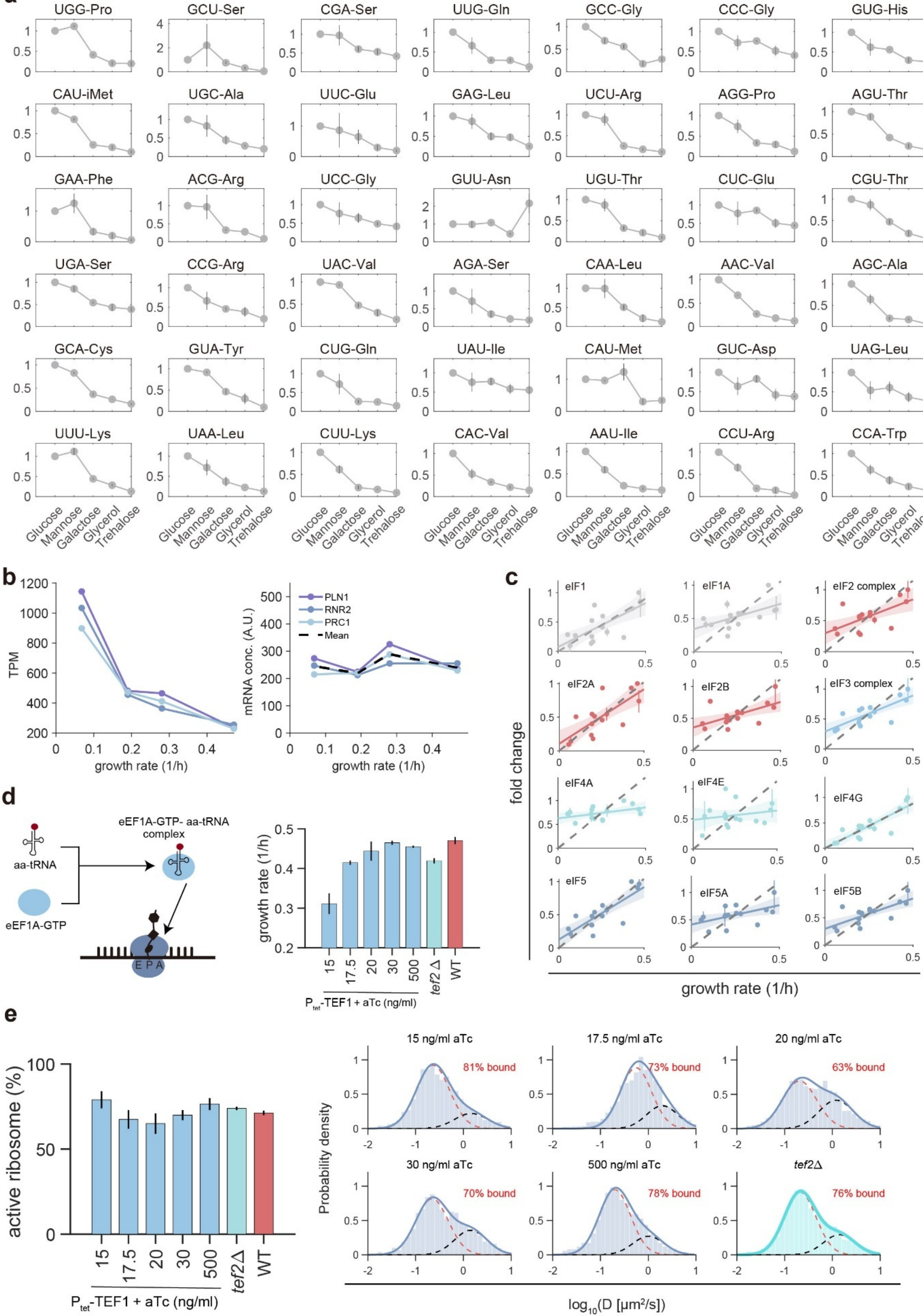

**Extended Data Fig. 4: Analysis of tRNA concentrations and eEF1A downregulation, related to Fig. 3.**

**a**, Fold changes in concentration of all 42 tRNA species across the indicated conditions are shown. Data represent the mean and range (n = 2).

**b**, Expression analysis of housekeeping genes *PLN1*, *RNR2*, and *PRC1*. Transcripts per million (TPM, left panel) decrease slightly with increasing growth rate, while their mRNA concentrations remain largely constant across conditions. The relative mRNA concentration of a given gene is calculated as TPM multiplied by the global mRNA concentration (see **Fig. 4a**).

**c**, Fraction of the proteome plotted against growth rate for initiation factor eIF1, eEF1A, eIF2, eIF2A, eIF2B, eIF3, eIF4A, eIF4E, eIF4G, eIF5, eIF5A, and eIF5B. Solid lines indicate linear fits with 90% CI. The dashed black line denotes proportional scaling with growth rate. Data are presented as mean ± range (n = 3 for glucose; n = 2 for galactose, glycerol, and trehalose) and normalized to the maximal value.

**d**, The growth rates are shown for the $P_{tet}$-*TEF1 tef2Δ* strain treated with different aTc concentrations, the *tef2Δ* strain, and a wild type strain. Data are mean with SEM (n=3).

**e**, The fraction of active ribosomes are shown for the $P_{tet}$-*TEF1 tef2Δ* strain treated with different aTc concentrations, the *tef2Δ* strain, and a wild type strain. Data indicate the mean and range (n=2). For the diffusion tracks: n=2,867 with 15 ng/ml aTc; n=2,268 with 17.5 ng/ml aTc; n=4,342 with 20 ng/ml aTc; n=2,365 with 30 ng/ml aTc; n=2,814 with 500 ng/ml aTc; n=3273 in the *tef2*Δ strain.

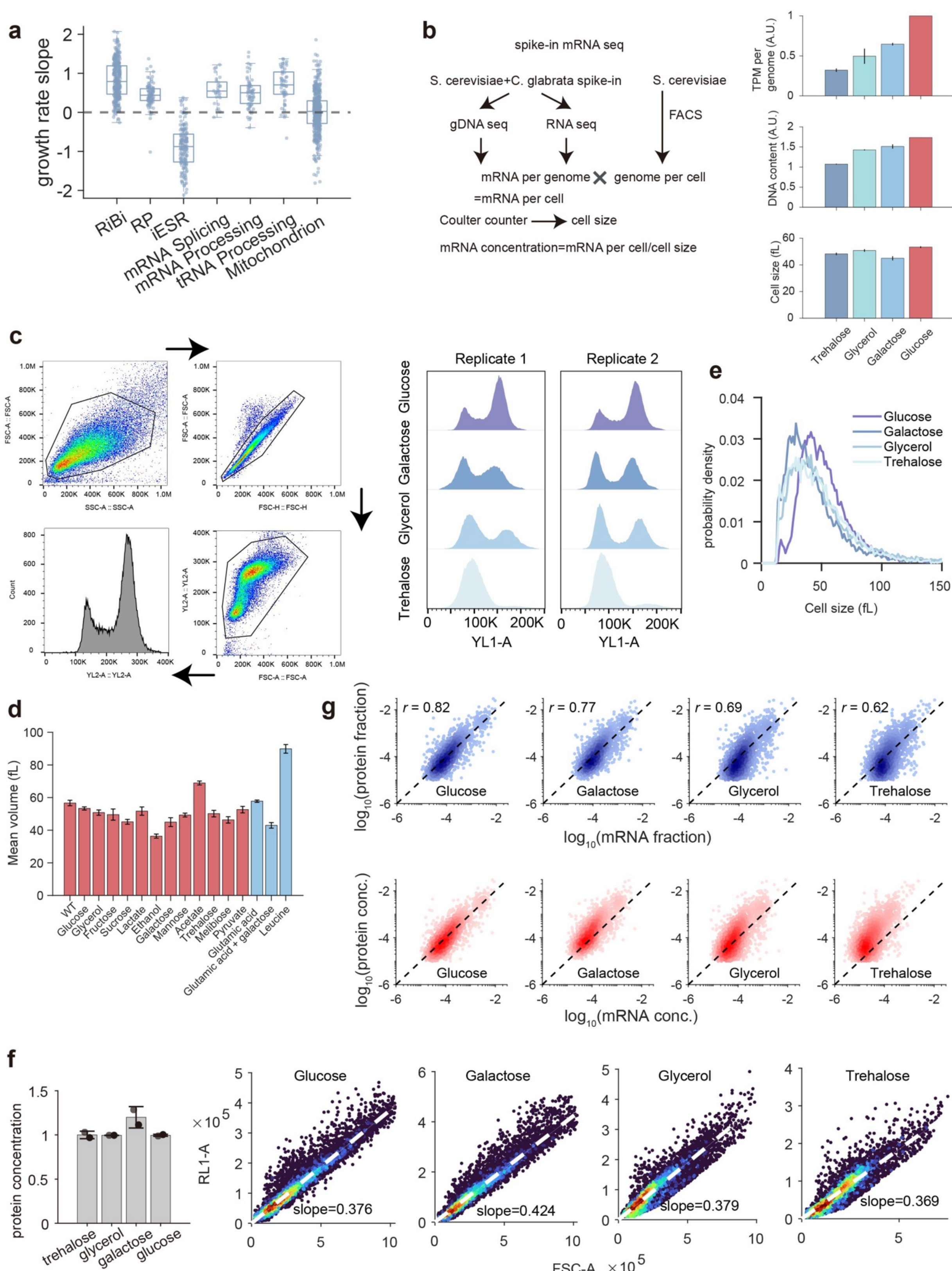

**Extended Data Fig. 5: Spike-in RNA sequencing to determine global mRNA concentrations, related to Fig.4.**

**a**, The box plot illustrates the slopes of protein abundance versus growth rate for ribosome biogenesis (RiBi), ribosomal proteins (RP), induced environmental stress response (iESR), mRNA splicing, mRNA processing, tRNA processing, and mitochondrion. Each box shows the median and interquartile range (IQR). The classification of these proteins is provided in **Supplementary Table 2**.

**b**, Workflow schematic of spike-in RNA sequencing to quantify global mRNA concentration. Right: normalized TPM per genome (n=2), average DNA content per cell (n=2), and average cell volume (n=3) measured under these four conditions. These parameters were used to calculate the global mRNA concentration. Data are mean with SEM.

**c**, Left panel shows the flow cytometry gating strategy. The right panels show DNA content distributions of cells cultured in SC medium with four different carbon sources.

**d**, Average cell volume (mean ± SD, n=3) for yeast populations cultured in SC media with 12 different carbon sources and 3 nitrogen sources.

**e**, Characteristic cell size distributions from Coulter counter measurements of cells growing in the indicated conditions.

**f**, Global protein concentration measurements by flow cytometry. Cells were stained using NHS-ester dye, and protein levels were quantified from the slope of robust linear regression of fluorescence intensity against forward scatter (FSC-A) shown in the right hand panels. Data are presented as the mean and range (n = 2).

**g**, Correlation of protein and mRNA fractions and concentrations for 1,795 genes across four conditions with different growth rates. Pearson correlation coefficient *r* was shown in the panel. Proteins with fractions < $1\times10^{5}$ were filtered out.

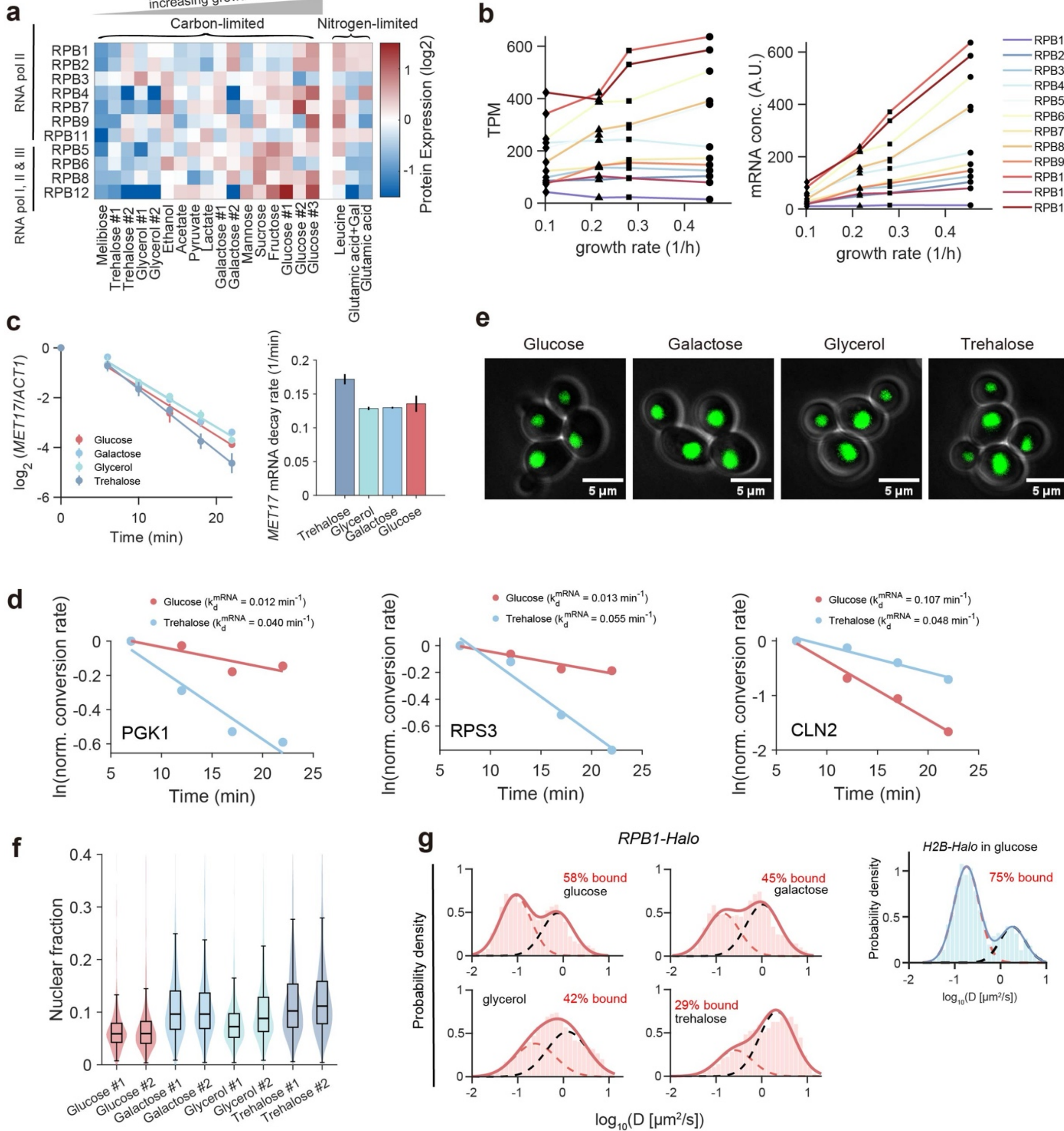

**Extended Data Fig. 6: Analysis of RNAP II activity, related to Fig. 4**

**a**, Protein concentrations for the 12 RNA polymerase II subunits across12 different carbon sources.

**b**, TPM and mRNA concentrations for the 12 RNA polymerase II subunits across12 different carbon sources.

**c**, Left: *MET17* mRNA concentration relative to *ACT1* mRNA concentration after methionine addition to the media for cells growing in the indicated carbon sources. Right: mean (± range) *MET17* mRNA decay rates (n = 2).

**d**, Normalized T-to-C conversion fractions over time for three representative mRNAs in glucose and trehalose. Lines indicate exponential fits used to estimate decay rates.

**e**, Representative composite phase and fluorescence images of Pus1-eGFP expressing cells growing in SC media with the indicated carbon source.

**f**, Violin plots showing the distribution of nuclear volume fractions for cells growing in SC media with the indicated carbon sources (n=687 for glucose #1, n=1,054 for glucose #2, n=503 for galactose #1, n=886 for galactose #2, n=475 for glycerol #1, n=1,286 for glycerol #2, n=515 for trehalose #1, and n=1,430 for trehalose #2). Each plot includes a box indicating the median and interquartile range.

**g**, Distribution of diffusion coefficients for RNAP II, where the Rpb1 subunit was tagged with Halo (n=3944 tracks in glucose, n=5392 tracks in galactose, n=4581 tracks in glycerol and n=12786 tracks in trehalose). Right: Distribution of diffusion coefficients for histones, where the histone 2 subunit H2B was tagged with Halo (n=1,237 tracks).

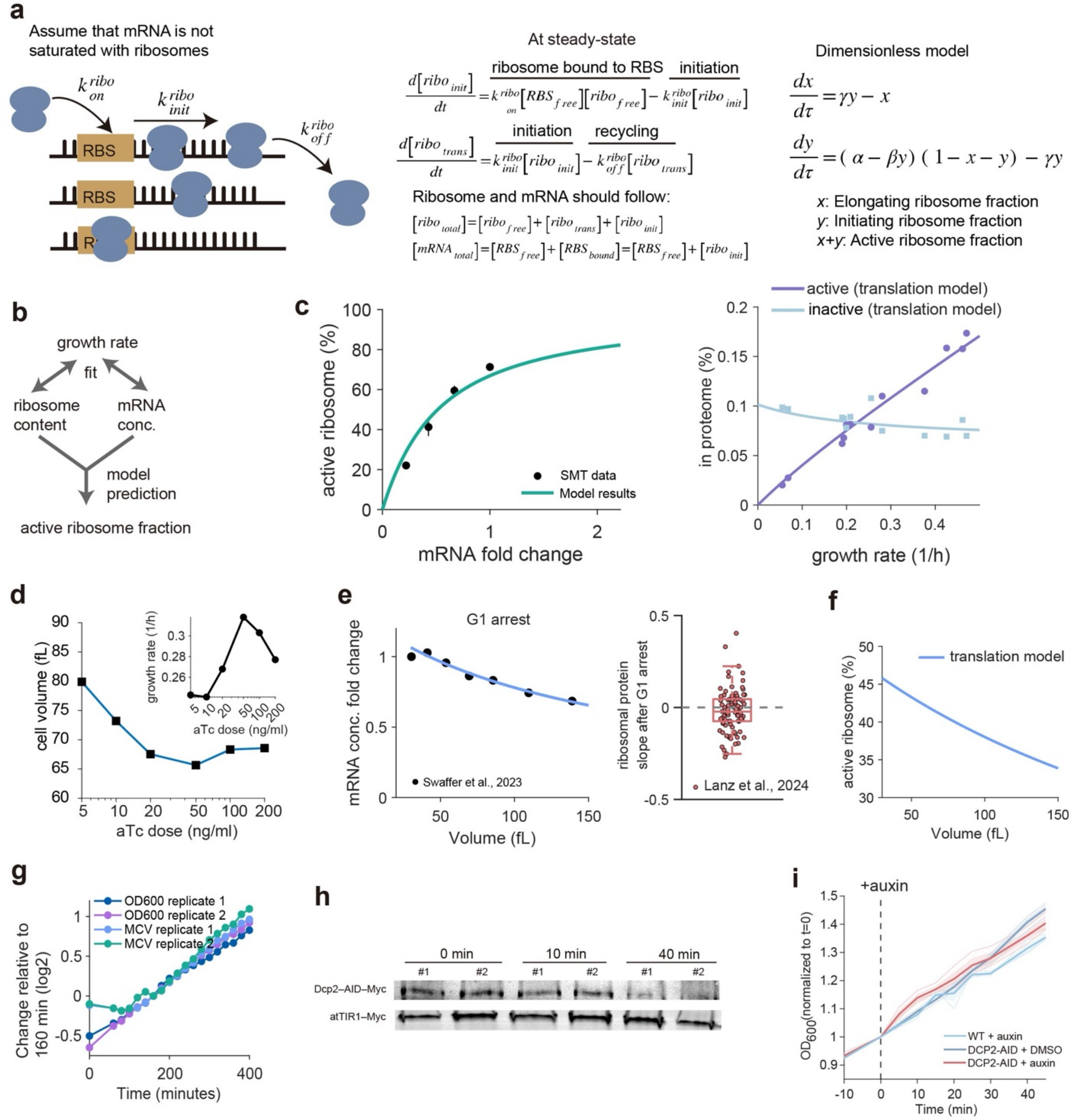

**Extended Data Fig. 7: A dynamic equilibrium model for protein translation, related to Fig. 5&6.**

**a**, Schematic illustrating the full dynamic equilibrium model, assuming ribosomes bind to mRNA without saturation. The system can be simplified into a dimensionless two-variable ordinary differential equation (ODE) model characterized by three parameters: relative mRNA concentration, relative ribosome concentration, and relative initiation rate (see **Methods**).

**b**, Workflow showing how we fit data to calculate the active ribosome fraction for each condition.

**c**, Comparison of data and the single parameter model fit predicting the fraction of active and inactive ribosomes.

**d**, Cell volume and growth rates at steady state for *$P_{tet}$-RPB1* cells growing in media with the indicated aTc concentration.

**e**, Left: After G1 arrest, the mRNA concentration decreases while cell volume increases. Data are from Swaffer et al [25]. Right: The slopes for fits of the normalized protein concentration versus normalized cell size for all the ribosomal protein subunits (for a detailed description of how these slopes were calculated, see Lanz et al [58]). A slope = 0 indicates a constant, size-independent concentration. Data are shown as median with IQR.

**f**, Dynamic equilibrium model prediction for the active ribosome fraction after G1 arrest.

**g**, OD600 and mean cell volume increase after G1 arrest. Data are normalized to values at 160 min. Measurements performed with two biological replicates.

**h**, Immunoblot indicating Dcp2-AID-Myc following auxin addition at indicated times. atTIR1-Myc served as the control.

**i**, OD600 time courses for DCP2-AID + auxin (n = 3), DCP2-AID + DMSO (n = 3), and WT + auxin (n = 3).